\documentclass[journal]{vgtc}                 % review
\ifpdf%                                % if we use pdflatex
  \pdfoutput=1\relax                   % create PDFs from pdfLaTeX
  \usepackage{graphicx}                % allow us to embed graphics files
  \DeclareGraphicsExtensions{.pdf,.png,.jpg,.jpeg} % for pdflatex we expect .pdf, .png, or .jpg files
\else%                                 % else we use pure latex
  \usepackage{graphicx}                % allow us to embed graphics files
  \DeclareGraphicsExtensions{.eps}     % for pure latex we expect eps files
\fi%

\usepackage{microtype}                 % use micro-typography (slightly more compact, better to read)
\PassOptionsToPackage{warn}{textcomp}  % to address font issues with \textrightarrow
\usepackage{textcomp}                  % use better special symbols
\usepackage{mathptmx}                  % use matching math font
\usepackage{times}                     % we use Times as the main font
\usepackage{cite}                      % needed to automatically sort the references
\usepackage{tabu}                      % only used for the table example
\usepackage{booktabs}                  % only used for the table example
\usepackage{pbox}
\usepackage{soul} % this is needed for the del command (sout)
\usepackage{xcolor}
\usepackage{listings}
\usepackage{caption} 
\usepackage[normalem]{ulem}
 
\definecolor{dorange}{RGB}{179, 83, 0}
\definecolor{pink}{RGB}{240,147,195}
\usepackage{tikz}
\DeclareRobustCommand*\circled[1]{\tikz[baseline=(char.base)]{
            \node[shape=circle,draw=pink,inner sep=1pt,fill=pink,text=white,draw=none, scale=0.8] (char) {#1};}}
\DeclareRobustCommand*\circledgray[1]{\tikz[baseline=(char.base)]{
            \node[shape=circle,draw=pink,inner sep=1pt,fill=lightgray,text=white,draw=gray, scale=0.8] (char) {#1};}}

\definecolor{emerald}{RGB}{106, 174, 169}
\definecolor{steelblue}{RGB}{70, 108, 155}
\definecolor{carrot}{RGB}{240,122,43}
\definecolor{crimson}{RGB}{220,77,79}
\setul{0.5pt}{0.8pt}

\newif\ifnotes
\notesfalse
\definecolor{purplecolor}{RGB}{128, 0, 128}
\definecolor{grey}{RGB}{128, 128, 128}

\DeclareRobustCommand{\hsout}[1]{\texorpdfstring{\renewcommand{\cite}{\ccite}\sout{#1}}{#1}}
\DeclareRobustCommand{\del}[1]{{\ifnotes{\leavevmode\color{grey}{\protect\hsout{#1}}}\fi}}
\newcommand{\add}[1]{\ifnotes{\color{purplecolor}{#1}}\else{#1}\fi}
\newcommand{\replace}[2]{\ifnotes{\del{#1}\add{#2}}\else{#2}\fi}

\usepackage{xspace}
\newcommand{\sys}{\textsc{diel}\xspace}
\newcommand{\diel}{\textsc{diel}\xspace}
\newcommand{\dcode}[1]{\texttt{\small\color{blue}{#1}}}
\newcommand{\apicode}[1]{\texttt{\small\color{dorange}{#1}}}
\newcommand{\smcode}[1]{\texttt{\small{#1}}}

\onlineid{0}

\vgtccategory{Research}

\vgtcinsertpkg

\title{DIEL: Interactive Visualization Beyond the Here and Now}

\author{Yifan Wu, Remco Chang, Joseph M. Hellerstein, Arvind Satyanarayan, and Eugene Wu}
\authorfooter{
\item
Yifan Wu and Joseph M. Hellerstein are with the University of California at Berkeley. E-mail: yifanwu@berkeley.edu.
\item Remco Chang is with Tufts University. E-mail: remco@cs.tufts.edu.
\item Arvind Satyanarayan is with MIT CSAIL. E-mail: arvindsatya@mit.edu.
\item
Eugene Wu is with Columbia University. E-mail: ewu@cs.columbia.edu.
}

\shortauthortitle{Wu \MakeLowercase{\textit{et al.}}: DIEL: Interactive Visualization Beyond the Here and Now}
\abstract{
Interactive visualization design and research have primarily focused on % operating with 
local data and synchronous events.
However, for more complex use cases---e.g., remote database access and streaming data sources---developers must grapple with distributed data and asynchronous events.
Currently, constructing these use cases is difficult and time-consuming; developers are forced to operationally program low-level details like asynchronous database querying and reactive event handling. This approach is in stark contrast to modern methods for browser-based interactive visualization, which feature high-level declarative specifications.
In response, we present \sys, a declarative framework that supports asynchronous events over distributed data. 
As in many declarative languages, \sys developers  specify only what data they want, rather than procedural steps for how to assemble it. Uniquely, \sys models asynchronous events (e.g., user interactions, server responses) as streams of data that are captured in event logs.
To specify the state of a visualization at any time, developers write declarative queries over the data and event logs; \sys compiles and optimizes a corresponding dataflow graph, and automatically generates necessary low-level distributed systems details.
We demonstrate \sys's performance and expressivity through example interactive visualizations that make diverse use of remote data and asynchronous events.
We further evaluate \sys's usability using the Cognitive Dimensions of Notations framework, revealing wins such as ease of change, and compromises such as premature commitments.
}

\begin{document}


\firstsection{Introduction}
\label{sec:intro}

\maketitle

% note: need to remove the following section if we are using tvcg template
% \section{Introduction}
% \firstsection{Introduction}

% Assert that prior work was too client-centric, and point out the challenges that they face when they need to work with remote stuff
Recent advances have made authoring browser-based interactive visualizations quite simple, via novel abstractions for specifying encodings~\cite{wilkinson2006grammar, bostock2011d3}, layout, data transformations, and interactions~\cite{satyanarayan2014declarative, satyanarayan2017vega}.
% why declaratively is great
Critically, these abstractions enable \emph{declarative} specification: developers can compose these visualization elements and defer execution concerns to the toolkit runtime.  
Declarative abstractions thus enable developers to build and iterate on designs quickly and expressively, while reducing the errors that would arise from imperative implementation~\cite{heer2010declarative, satyanarayan2014declarative}.
% issue
However, these abstractions do not gracefully support use cases where data may be distributed across multiple locations, or where events cannot be synchronously handled to completion.
% example
To illustrate, we discuss two interactive visualization settings---one quite traditional and another more complex---and reflect on the difficulty of addressing these issues with current practices.

% A: This paragraph feels too in the weeds. It's focused on describing _how_ the developer would set things up because the abstractions don't exist, rather than the impact/implications. What are the interesting conceptual issues these examples highlight? i.e., what kind of activity can the developer not easily engage in because the abstractions do not exist? or what difficulties do they experience?
Consider the \emph{client-server architecture}, a classical design for scalable interactive visualizations that remains common even in recent research~\cite{moritz2019falcon, moritz2017trust}. A concrete example is shown in Fig.~\ref{fig:space_time_examples}\circled{1}, where the brush selection over the scatterplot is computed by a remote database.
To handle the brush selection, a developer can no longer rely on existing declarative abstractions (e.g., Vega-Lite's selections~\cite{satyanarayan2017vega}) alone.  Instead, they must now translate interaction logic from a visualization specification to the query language of the database, network with the remote database, and manually bookkeep the provenance of interactions and database responses to ensure that the interface is displaying the correct results, in the correct order, in the face of non-deterministic processing latency. 

% These additional instrumentation is not limited to interactive visualizations over large datasets.  
A more specialized example is \emph{streaming data}, increasingly investigated in research~\cite{xie2007towards, krstajic2013visualization, wanner2014state} and industry~\cite{zoomdata}.
An example is shown in Fig.~\ref{fig:space_time_examples}\circled{2}, where the points can be streamed in the scatterplot and selected by a brush interaction.  
The two processes\,---\,the user’s brushing interaction, and the tweet stream\,---\,introduce ambiguity about what should happen when both attempt to update the visualization concurrently.
For instance, consider the case when a new point streams in, falling within a brushed region.  Should this new point be part of the existing selection, or should it be isolated in a new separate selection? Should the new point be blocked until the user removes the brush selection, or should the point be added and the brush selection removed?
These decisions can be further complicated if the brush selection is processed by a remote server, which introduces another asynchronous event that multiplies the complexity.

% Yet another type of examples include \emph{Scented Widgets}~\cite{willett2007scented}, \emph{HindSight}~\cite{feng2017hindsight}, and \emph{Interaction Snapshots}~\cite{wu2020snapshot}.  These designs require past interactions and their temporal information to be made available.  With current frameworks, a developer would need to resort to manual bookkeeping.

% consequence
For a designer to explore decision space fluidly~\cite{green2000instructions}, we want to to minimize the friction of experimenting with these choices. But distributed and streaming visualizations still have a high programming barrier, despite progress on declarative libraries for browser-based data visualizations. The effect is that designers get ``cornered'' into whatever is easiest: for example opting for a high-latency blocking design, or disallowing users from interacting with streaming data, purely for ease of implementation rather than superior usability.
% , where users cannot interact until the system has responded,
% sign post
To improve design fluidity in these use cases, we identify two key abstractions that are missing from existing interactive visualization frameworks:

\begin{figure}[bt]
 \centering
 \includegraphics[width=0.9\linewidth]{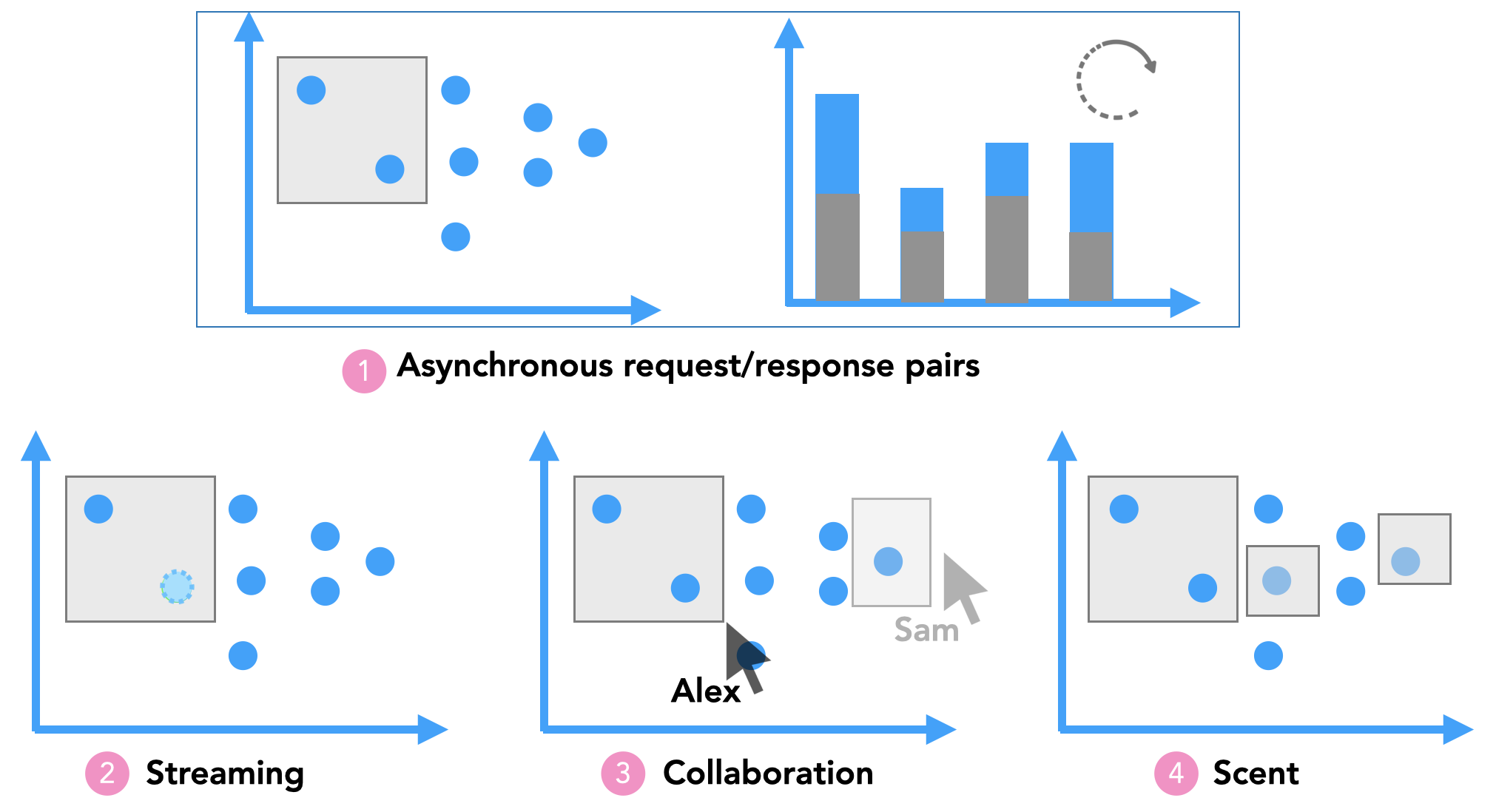}
 \caption{\replace{Example interactive visualization use cases where data may be remote and/or events are evaluated asynchronously.}{Example interactive visualization use cases where (1) data may not fit into the browser and thus lives in a remote database, and/or (2) events, including user interactions and data updates from servers, need to be evaluated asynchronously.}  \add{Concretely, \circled{1} shows interactions taking time to process, creating asynchrony between interactions and result updates (see Fig.~\ref{fig:async_policy} for details). \circled{2} shows data updating in real time as the user is interacting with a visualization. \circled{3} shows an example collaboration experience~\cite{neogy2020representing}; \circled{4} shows a visualization of past interactions to curate ``information scent'' (see Fig.~\ref{fig:scent}).}}
 \label{fig:space_time_examples}
\end{figure}

% First, it may not be possible to compute the data synchronously on the main browser thread as the data needed for the application may be distributed in remote databases.
% Second, propagating events through the system in a transient manner is insufficient as the history of prior events may impact events in the future. 
% Data is increasingly distributed across different compute nodes.   
\noindent\textbf{Distributed Data.}  Data is increasingly distributed at different compute nodes\del{ (e.g., across a browser and a remote database)}
% {, which could be across multiple servers hosting a large dataset, or, in the case of a user facing application, across browser threads and a remote database}.
\add{Traditionally, this was across multiple database servers hosting a large dataset, however user-facing applications increasingly distribute data across browser threads, and across browser state and remote databases as well.}
As such, developers currently need to write code to access the data, which they often need to further optimize. Our goal is to develop abstractions that (1) free developers from sending data manually between local and remote processes, (2) allow developers to easily change their data storage and compute sources (e.g., between the main browser thread, a WebWorker~\cite{webworker}, a local database, or cloud databases) without rewriting their application code, and (3) unobtrusively perform low-level optimization (e.g., caching).

% :)
\noindent\textbf{Asynchronous Events.}  When data is local on one client, user interactions are the only events that a developer has to work with. Thus, computation caused by interactions, such as filtering data based on a selection, can be handled synchronously. However, when data is remote, asynchronous events are the norm: responses from a server, streaming data from real-world events, or events generated by other users. These asynchronous events are outside of the application's direct control and, thus, have to be handled as they arrive.
% us
Our goal is to develop abstractions \add{and a library} that allow developers to (1) create consistent user experiences in the face of concurrency and out-of-order execution, and (2) easily experiment with different designs for handling asynchronous events.
% , such as blocking and non-blocking interfaces.
% There are many other types of temporal designs, which we show in Section~\ref{sec:examples}.

To effectively work in this new paradigm of distributed data and asynchronous events, we adapt two patterns from the field of distributed systems programming:
% we need to further rethink core assumptions and make use of abstractions designed for distribution and asynchrony.   
% To develop a model and a runtime to meet these two requirements, w
% One field that specializes in these challenges is distributed systems programming.
% , which specializes in dealing with distributed data and asynchronous events.
%  where developers work with data and computation spanning multiple locations and non-deterministic asynchrony.
% Two patterns have served the field well.

\noindent\textbf{Logical Constraints.} Rather than imperatively manipulating data, developers express logical constraints (i.e., queries) over data, and the system determines the methods used to execute the queries~\cite{codd1970relational, ramakrishnan2000database}.  By decoupling the \emph{how} from the \emph{what}, logical constraints allow the system to plan and optimize the execution of the query (possibly spanning multiple databases)~\cite{ramakrishnan2000database}, relieving the developer of this burden.
% literature from \emph{federated databases} and query optimization techniques such as materialized view maintenance.
% Furthermore, languages based on relational algebra, such as SQL and dataframe libraries, which express logical constraints over data, are widely supported by various databases that data visualization practitioners are already using.
This approach also has practical implications. In particular, SQL and dataframe libraries express logical constraints over data, and have been widely adopted by developers and database engines~\cite{sqlite, postgresql, armbrust2015spark} alike.
% \footnote{Some dataframe libraries such as \texttt{pandas} offer more than relational operators but all support relational algebra such as select, project, and group by~\cite{petersohn2020towards, wu2020dataframe}} 
% Not only does are relational queries easy to specify, they can also be optimized by the system, rather than the developers, who are users of the system---there are a wealth of literature and implemented techniques that help with  optimization.

\noindent\textbf{Immutable Events.}  Past events, along with their time steps, are stored as an immutable log by the system.  Developers declare the current state of the application as a logical constraint over the immutable log rather than transiently handling events and mutating state via side effects~\cite{helland2015immutability, alvaro2011consistency}.  
As a result, developers can coordinate events declaratively without worrying about how to update the application state through callbacks over transient events.
% The system provides developers with access to the atomic timesteps to coordinate asynchronous events.
% Another benefit is that developers can work with timestamped records instead of .  They 
% The system takes care of executing the distributed dataflow.

In this work, we bring these ideas from distributed systems programming to the context of interactive visualization.
\sys\footnote{\textbf{D}eclarative \textbf{I}nteraction \textbf{E}vent \textbf{L}og} models interactions and asynchronous data computation events as timestamped records in event tables, and the state of the interface as a query over the event tables and data tables. 
The \sys runtime maintains an event loop to reevaluate queries as event tables change, ensuring that the interface is responsive to interactions.
% consequence
With \sys, we are able to achieve the requirements outlined earlier\,---\,developers no longer need to implement low-level networking with remote databases or perform optimizations, and have a way to easily coordinate asynchronous events that maintain a consistent interface.

To evaluate \sys's expressiveness, we construct a diverse set of interactive visualizations over distributed data and asynchronous time. Examples include ways to handle request-response asynchrony~\cite{wu2016devil, wu2020snapshot}, interactions on streaming data, composing two related interactions, and interaction scents~\cite{willett2007scented, feng2017hindsight}.
These examples show that \sys's abstractions allow developers to rapidly and concisely explore different interface designs.
% by allowing developers to express the interface state as logical constraints over event and data tables, \sys takes care of working with distributed data and promotes declarative design of ways to work with asynchronous events.  We also show that
We also present a heuristic analysis of the usability of these abstractions using the \emph{Cognitive Dimensions of Notation} framework~\cite{blackwell2001cognitive}, highlighting both gains to fluidity, and compromises to premature commitment. 
Finally, we verify the viability of \sys through performance measurements of a prototype, and show that \sys adds low overhead and scales as data increases in size.

\section{Related Work}

\sys builds on prior work in databases, visualization systems, and distributed systems programming.

%%%%%%%%%%%%%%%%%%%%%
% \red{joe workshop to finish}
\noindent\textbf{Database and Visualization Systems.}
% Relational abstractions have proven to be a powerful and enduring API for working with data. (Maybe say something about the reemergence of SQL for big data etc.) They're good because usual stuff we already abour relational model. Also, the techniques for parallelizing and distributing SQL across multiple repositories are well understood... something we will exploit in Sec~blah.
% merging
Research at the intersection of visualization and database systems has traditionally focused on enhancing performance\,---\,for instance, by offloading data aggregation and filtering to a remote database~\cite{stolte2002polaris, moritz2017trust} or by embedding a high-performance database-like query engine into the browser~\cite{satyanarayan2017vega}.
% There are two common architectures.
% One way to improve performance is to offload data aggregation and filtering from the front-end into the backend database~\cite{stolte2002polaris, moritz2017trust}.
% Another is to embed a high-performance database-like query engine into the browser~\cite{satyanarayan2017vega}.
In contrast, our work is \emph{not} motivated primarily by performance but instead by the  programming experience of visualization developers.
\del{In particular, \sys should be thought of as a programming abstraction between a database and a client.} 
Although \sys embeds a SQL query engine in the browser, it does so for its benefits as a \emph{programming construct}, and its ability to interoperate with a remote database.
Critically, \sys is agnostic to the specific database system, which could range from a SQLite instance in the browser's main thread~\cite{sqljs}, to commonly used databases on a server, such as SQLite, Postgres~\cite{postgresql}, or new research systems~\cite{liu2013immens,psallidas2018smoke}.

% \ewu{Would try to rephrase.  Something like: 
FORWARD is a general web application programming framework that directly binds database records and query results to DOM elements~\cite{fu2014forward}.  \diel is similar in that it manages how elements in the visualization are updated based on changes in the underlying database that represents user interactions.  However, \diel uniquely addresses the challenges of asynchronous events, which are shown to be especially challenging for interactive visualizations~\cite{wu2016devil}.  \sys does so through a novel declarative programming API over event histories.
\noindent\textbf{Visual Programming over Tables.}  There is a close connection between visual querying and textual queries\,---\,visual querying systems combine textual query specification with direct manipulation and data visualization.
For instance, VIQING maps visual selections, joins, and reordering to SQL operators~\cite{olston1998viqing}.
% Similarly, DEVise designed the abstraction of \emph{cursors} and \emph{links} between queries~\cite{livny1997devise} to express complex cross-linking logic.
Polaris endowed the pivot operator with powerful interactive capabilities~\cite{stolte2002polaris}.
Wrangler brings together interactions and a query based DSL for data transformations~\cite{kandel2011wrangler}.
\add{Many commercial tools, such as Trifacta, Tableau, and Microsoft Power BI/Query, provide a combination of these capabilities.}

\add{Beyond the connection between queries and interactive visualization, Psallidas and Wu demonstrated the application of query \emph{lineage} in interactive visualizations~\cite{Psallidas2018ProvenanceFI}.} \replace{B2 leverages query lineage to instrument}{Also leveraging lineage, B2 instruments} interactive cross-filtering visualizations in computational notebooks~\cite{wu2020b2}.
% similarities
\sys builds on these close connections using relational queries to define the data transformation logic of interactions.
However, \sys differs from these systems in that it is a programming abstraction \replace{and}{for} a middleware layer \replace{for}{designed to support} general purpose interactive visualization programming.  
% Developers use \sys in conjunction with front-end visualization libraries to support general purpose interactive visualizations.

%%%%%%%%%%%%%%%%%%%%%
\noindent\textbf{Interactive Visualization Libraries.} While \sys makes use of interactive visualization libraries, such as D3~\cite{bostock2011d3} and Vega~\cite{satyanarayan2017vega}, it is not one itself.  \sys relies on the front-end visualization libraries to map data to visual encodings, perform visualization-specific transformations (such as \smcode{voron\"{i}}, \smcode{treemap}, or \smcode{wordcloud}~\cite{vegaTransform}), create interactors, and capture selection values.
% gap \sys fills
Existing abstractions in these libraries support interactions in a \emph{synchronous} setting where the computation is expected to finish immediately. As a result, the events are also transient by default and assumed to be unnecessary for subsequent events.
In contrast, \sys supplements existing frontend visualization libraries by filling in the gap of working with distributed data and asynchronous events.
Of particular note is how \sys differs from Vega, which also models interaction events as streaming data.
% \joe{Reactive?}
Like other libraries, events in Vega are transient by default and, although a rich set of data transformations are offered, they only operate over client-side data.
\sys, on the other hand, records all events into a persistent log and offers a simple set of relational abstractions over distributed data.
% For instance, events in Vega are transient by default, whereas \sys records all events into a persistent log.
% Similarly, Vega supports a rich set of transformations but only for local data, while \sys supports simple relational abstractions over distributed data.
Future research could consider how to further extend Vega's dataflow abstractions~\cite{satyanarayan2014declarative, satyanarayan2016reactive} to better integrate with, or adopt \sys's abstractions.
% It is possible to extend Vega's dataflow abstractions~\cite{satyanarayan2014declarative, satyanarayan2016reactive} to work with that of \sys's, which is not trivial and merits further research.

%%%%%%%%%%%%%%%%%%%%%
\noindent\textbf{Frameworks for Provenance and Asynchrony.}
With current programming paradigms, developers need additional instrumentation to support features like logging or undo/redo.
For example, \emph{Trrack} is a purpose-built library that augments existing interactive visualizations with a provenance graph of state history~\cite{cutler2020trrack}.
% It provides a flexible API for existing applications.
% that works with arbitrary state, making it better suited for
%  does not augment existing applications, but is
\sys, in contrast, offers a more general-purpose set of abstractions which provide provenance tracking via first-class event histories.
% In contrast, \sys is a new way to build applications which provides provenance tracking via first-class event histories.

%%%%%%%%%%%%%%%%%%%%%
% \noindent\textbf{Frameworks for Asynchrony} 
Similarly, to be able to share and synchronize multiple users' changes over the network, programmers of collaborative groupware applications rely on special-purpose frameworks.
% (CSCW) community has proposed solutions to program  where .  
Toolkits such as \emph{Janus} help developers resolve concurrent and possibly conflicting events to ensure that each participant views a consistent artifact~\cite{savery2011s}.
\sys takes inspiration from these systems, e.g., the time-stamped events in Janus~\cite{savery2011s}, but also adds specialization for interactive visualizations, such as providing support for working with distributed data and tracking request-response provenance.
% another
\sys was directly inspired by distributed
programming language research, notably the Bloom~\cite{alvaro2011consistency} language and  CRDT~\cite{preguica2009commutative} data structures. 
Like Bloom, 
% (and in contrast to the eventually-consistent CRDTs)
% focuses on resolving the \emph{eventual} artifact, whereas 
\sys focuses on the semantics of atomic timesteps immediately after the user interacts,
% \sys makes timesteps atomic and explicit to 
helping developers reason about out-of-order events. \sys differs from the prior work in its focus on reifying a log of history as a core aspect of its data model.

%%%%%%%%%%%%%%%%%%%%%
\add{
\noindent\textbf{Reactive Programming.} 
\emph{Reactive Extensions} (Rx) is a widely-used example of a Functional Reactive Programming (FRP) library, which coordinates event-based and asynchronous computations such as mouse moves and high-latency calls to Web services~\cite{meijer2012your}.
% Among the languages we discussed, 
Vega~\cite{satyanarayan2014declarative} and Bloom~\cite{alvaro2011consistency} also follow the model of reactive programming.
\sys takes inspiration from these libraries, implementing reactive programming semantics.
% \sys takes many design inspirations from Rx.  
Meijer's article about Rx, \emph{Your Mouse is a Database}~\cite{meijer2012your}, informed our formulation of user events as record insertions into tables.

However, compared to Rx, \diel has a much more restrictive model targeted at visual analytic applications.  Unlike Rx, \diel does not support general purpose applications but provides domain specific functionalities for the use cases in interactive visualization.
One could implement \sys-like logic as a specialized design pattern in an FRP language like Rx but doing so would involve implementing the distributed query execution logic that \sys provides, which is not a casual task for a visualization developer.
}
%! TEX root = ../diel_eurovis.tex
% \section {Towards a Distributed and Asynchronous Model of Interactions}
\section{Programming Interactions Across Space \& Time}
\label{sec:rational}

To address the challenges of asynchronous events and distributed data, we look to distributed systems programming for inspiration and adapt the ideas to interactive visualization programming. 

%%%%%%%%%%%%%%%%%%%
\subsection{Asynchronous Interactions: Immutable Events}

Asynchronous events, unlike their synchronous counterparts, need to be coordinated by developers to ensure a good user experience~\cite{wu2016devil}.
% , as shown in examples from Fig.~\ref{fig:space_time_examples}.
This coordination can be challenging as  developers need to maintain relevant information of past events and selectively trigger event handlers.
And, since different parts of the event handling and bookkeeping logic are scattered across functions and variables, these imperative bookkeeping and callbacks can become tedious to maintain and prone to errors. 
This situation is exacerbated when new types of events need to be supported (e.g., a new interaction).  
To address these challenges, we look to key methods in distributed systems programming.

% The challenges are similar to the problems that  have to deal with, and 
\noindent\textbf{Events as Data.} An important technique in distributed systems programming is making events immutable and first class~\cite{helland2015immutability}.  Events are stored as a log, and the state after each event is defined as a function (or query) of the log~\cite{abiteboul2000relational}.
% say more about what's good
This design abstracts away the details of maintaining state (no more mutations with callbacks) and makes it easier to reason about the consistency of the application~\cite{alvaro2011consistency}.
% here
We can apply this framing directly to interactive visualizations: user interactions are events, and queries over these events (and other tables) can specify the state of the interface.
% As both the events and the underlying data are represented as tables, the output can be expressed by the same query syntax as over static input data.
% a \emph{relational transducer} evaluates the values of ``output'' views after the underlying ``input'' tables change.

\noindent\textbf{Atomic Timesteps.}  Compared to synchronous events, asynchronous events can, by definition, arrive out of order.
This behavior can lead to a variety of ``inconsistent'' visualization states, as documented by Wu et al.~\cite{wu2016devil}.
One simple example is naively rendering a response when it is received, even if it is for an ``old'' interaction made prior to the most recent one.
To deal with unpredictable sequences of events, developers need to reason about what prior events occurred, and when they did so.

% For example, one straightforward strategy is to synchronize events. 
Given this challenge of inconsistency, one obvious solution is to synchronize the events. 
However, this solution violates a core HCI principle: \emph{responsiveness}~\cite{hutchins1985direct, johnson2007gui}.  Making events synchronous means that the user is \emph{blocked} from performing new interactions until their previous interaction result has been executed completely (including being sent to the server, computed on the server database, received by the client, and rendered to the UI).
As this process can take a significant amount of time to complete, making the interface unresponsive yields a frustrating user experience.
% the user when they wish to perform another unrelated interaction or discard the previous result.

To maintain a smooth user experience, it is important to work \emph{with} asynchronous events. Here, one idea from distributed programming is particularly helpful: \emph{atomic timesteps}~\cite{alvaro2011consistency}, whereby the system guarantees an atomic unit of evaluation by computing the state of the application in full before admitting another input event.
With atomic timesteps, developers reason independently, ``frame by frame'', about what computation needs to be computed synchronously within a given timestep.
Moreover, developers can reference the explicit timesteps stamped on events, with the guarantee that an event with a smaller timestep precedes an event with a larger timestep.
Finally, developers can be sure that the interface satisfies the constraints specified at every timestep, which could support robust UX in the face of asynchrony.

% The system does not wait for for other asynchronous events, which could take a while, but waits for synchronous evaluations that are deemed important by the developer.
% and allow developers to access and reference the timesteps when defining the state of the system.
% This idea of timesteps applies directly to visualizations: common interfaces often display the result of the most recent user interaction event, which is the event with the highest timestep.

%%%%%%%%%%%%%%%%
\subsection{Distributed Data: Logical Constraints}

To be able to scale an interactive visualization application, a developer must be able to flexibly change where the data is stored and computed upon.
They may initially build a prototype over small datasets that are stored and computed in the main thread of the browser.  
% move the storage and compute to
To use a larger dataset, they may utilize WebWorkers~\cite{webworker}, which allow the data-rich tasks to run asynchronously in the background and not block the interface.
Then, as the developer deploys the application to real-world datasets, they may move the computation to a database on their local machine or a cloud database.
% For developers, it is important to be able to adapt their program for data of different sizes means that they should be able to use data in different locations, such as the browser, WebWorkers, or remote databases.

% goal
% \vspace{-2mm}
To achieve this flexibility, the system should abstract data access details from developers.
% db
The field of databases has tackled this problem using a concept that is not unfamiliar to interactive visualization developers: \emph{relations} (also called \emph{tables}). Relations are sets of data tuples with a fixed schema, and computation over relations is governed by an algebra of select, project, join, and group by operators~\cite{codd1970relational}.
Relational query languages bring with them two important properties for our purposes:

% to reason programmatically with the computation and
%  execute the computation to
% , which offers two key features for programming interactive visualizations.

%%%%%%%%%%
% \vspace{-2mm}
\noindent\textbf{Physical Data Independence.}  
% issue 
Currently, to change from storing and computing data in the main browser thread to any other options requires custom code: developers may need to map interaction logic from JavaScript to SQL, handle the database connection, and perform networking.
% why
This work is needed for two reasons.
First, developers often specify \emph{how} to access the data.  If the data location changes, the program changes.
Second, the computation abstractions on the client and other locations may be different.
Relational languages can address both of the factors.
They allow developers to specify \emph{what} data to access, making the program independent from the physical details of the data.
Furthermore, if a standard relational language is used everywhere, be it the client, a WebWorker thread, or a remote database, the developer can work with one abstraction and not have to translate between specifications.

% map the logic from in JavaScript to SQL, and could instead rely on a single, shared abstraction.

%%%%%%%%%%%%%
% \vspace{-2mm}
\noindent\textbf{Rich Optimization}  Currently, developers may implement optimizations manually, such as caching. 
This is not ideal for two reasons: one is a higher programming barrier, and another is that the developer may not have enough time for more involved optimizations beyond a simple cache.
% how RA helps
\replace{Relational languages}{Having relational abstractions over both the client and the server can} relieve the optimization burden from the developer and allow the system to compile the logical specification into a physical execution plan, using a wealth of optimization techniques~\cite{ramakrishnan2000database}.

% When implementing client-server architectures for interactive visualization, the responsibility for optimizing processing falls to the developer as well. For instance, to avoid expensive query round-trips, a developer would need to implement a cache of previous interaction results. 
% Another example is to load relevant data from the server before allowing the user to interact, which front-loads the wait time to reduce interactive latency.

\section{The DIEL Model}
\label{sec:model}

\begin{figure*}[tb]
  \centering
  \includegraphics[width=\linewidth]{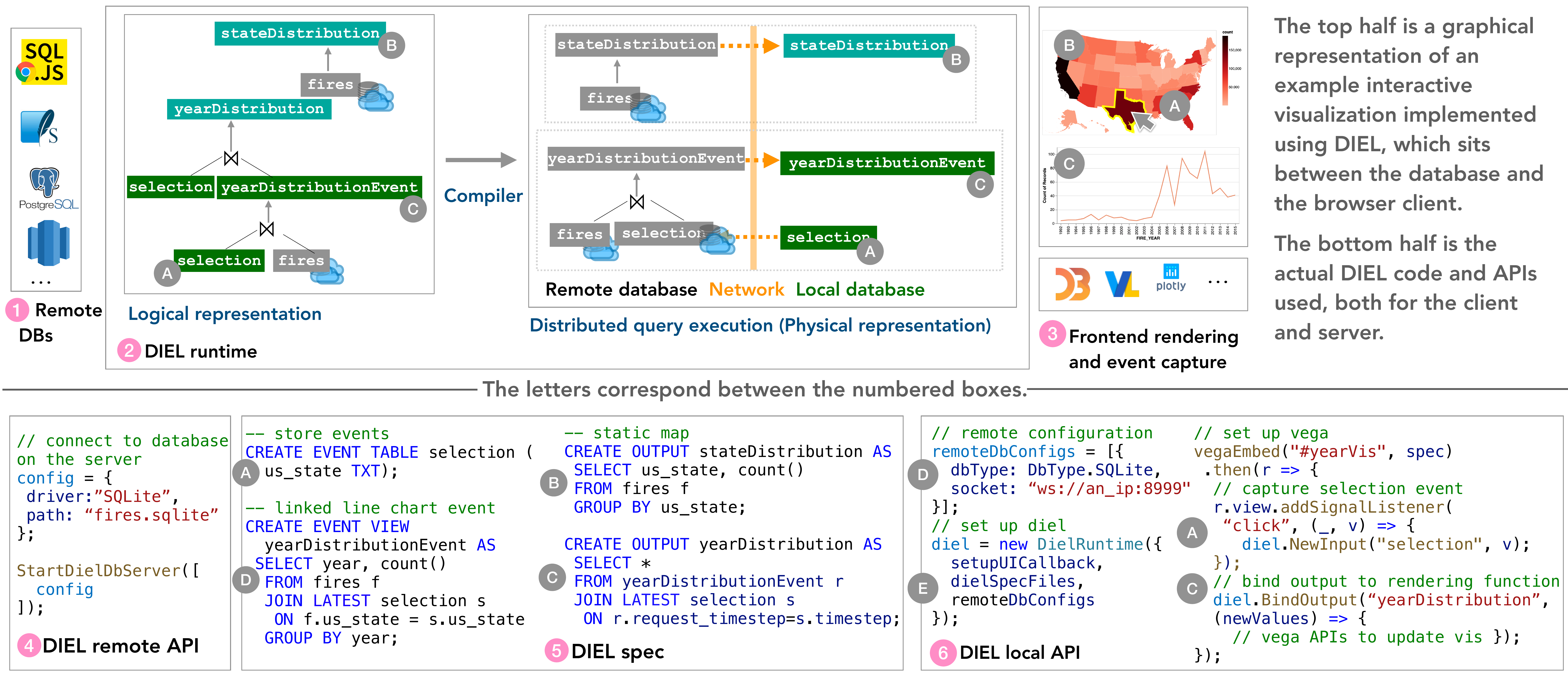}
  \caption{Example code using \sys to power the interactive visualization of wild fires in the US.  The user can click on a US state to filter the distribution of fires over the years \circled{3}.  To create this visualization, developers use \sys in conjunction with data visualization libraries and remote databases \circled{1}.
  Developers query over timestamped event logs (\smcode{selection}) and base dataset (\smcode{fire}) to specify the state of the application (a logical spec \circled{5}), and \sys orchestrates the computation between the local and remote databases (an execution plan \circled{2}).}
	\label{fig:example}
\end{figure*}

To address the challenges in Section 1, \sys manages user interactions, remote databases, and the communication between local and remote.
On the browser frontend, the developer specifies the data that the user can interact with and select (e.g., using Vega-Lite~\cite{satyanarayan2017vega}) and translates them into events that are sent to \sys via its JavaScript API.
\sys stores events in the local in-browser local database and manages query processing on the remote databases, which are connected to \sys through a set of remote-side APIs.  Through standard database connection libraries such as \smcode{node-postgres} and \smcode{sqlite3}, \sys allows the developer to connect to remote databases ranging from SQLite, to PostgreSQL, to cloud databases like Amazon Redshift.

Note that remote databases and their asynchronous complexities can arise in surprising settings.  In addition to databases on remote servers (such as Redshift), remote databases also arise when data is managed by a different process on the user's computer.  For instance, browser WebSockets communicate with the web page's main thread via asynchronous messaging.  For this reason, \diel is designed to work for any distributed database setting, irrespective of whether the databases are on the cloud.

The developer provides a specification that \sys uses to coordinate query processing between the local and 
% \ewu{The previous paragraph emphasized that "serverside" databases are needed in DIEL, so reader could be confused about "backend" database.  Maybe clarify?} 
remote databases.  The specification defines the application state that \sys will manage, and is written in a SQL-like language. We chose SQL as the basis because it is the most widely used data processing languages today~\cite{sosurvey}, and is familiar to developers working with large datasets.  However, \sys is not tied to SQL, and alternative relational languages. It is straightforward to layer a dataframe-like API atop \sys's abstractions.
Given the specification and runtime API calls, \diel coordinates the local and remote database to exchange data and query results. The query results then update visualization state back to the application via a JavaScript API so that it can render results and update the visualizations.

We next discuss how developers specify different types of application state, and how it is managed during run time.   Fig.~\ref{fig:example} depicts the running example, and Tables~\ref{tab:relations} and~\ref{tab:api} summarize the syntax.

\subsection{Data Model}

\noindent\textbf{Static Visualizations}
Static visualization is a well-accepted domain, where tabular data is fed to APIs that implement a grammar of graphics~\cite{satyanarayan2017vega, wilkinson2006grammar}. For static visualization, ~\diel is responsible for preparing the data to be rendered, and invoking the visualization API when the data is ready.
The tables used to create visualizations could either be stored tables or the results of queries.
For instance, in Fig.~\ref{fig:example}, the map visualization maps the fields \texttt{\small (us\_state, count)} from the query result in (5B) to polygons and fill color in the visualization.
% eval
When the application is first loaded, \sys evaluates the initial queries to render the static visualizations. Using the \apicode{BindOutput} API (\replace{6D}{6C}), the developer registers a callback \emph{rendering function} that is called when the query results are updated.
% \gray{The developer binds the evaluated results with a function that renders the visualization (called a \emph{rendering function}) through the \sys \apicode{BindOutput} API.}
% Developers then take the data shared and renders the visualization, through .
The developer annotates the queries that the rendering function will access using the \dcode{OUTPUT} keyword.  The system \replace{will use}{uses} the information to inform the dataflow as well as instrument the \apicode{BindOutput} API.
% create the correspondinginstances.
% compile the definitions into a  Javascript API (THAT WILL BE USED IN SOME WAY?).

% At runtime, when the values of the \dcode{OUTPUT} tables change, the rendering function will automatically be invoked with the new values.
% As shown in Fig.~\ref{fig:example}(4D), the output \texttt{\small yearDist} is bound to the function that updates the line chart visualization in Vega.
% With the the \texttt{\small .change} API of Vega, the developer can now push the new \texttt{\color{blue} \small OUTPUT} values to Vega which updates the UI.

\noindent\textbf{Events} \sys stores events as timestamped tables.
When the developer defines visualization interactions, such as selections, the data records that are selected are ingested into \sys through the \apicode{NewEvent} API and stored in tables.
% \joe{noun?}
% (called \emph{points of interest})
% \dcode{EVENT TABLE}s.
% \ewu{are translated into events?} that populate the corresponding event table.   
% \gray{Developers specify the schema of the event tables based on the \emph{points of interest} selected by the interaction (following the abstraction for selections in Vega-Lite~\cite{satyanarayan2017vega}).}
For instance, the selection on the map (3A) is mapped via a Vega listener (6A) and stored in the \texttt{\small selection} table (5A), with a single column, \smcode{us\_state}, representing the state selected, e.g., ``CA''.
An event may be one or more rows.
At runtime, each new event is augmented with a \emph{logical timestep}, which we discuss in Section~\ref{sec:exec}
% execution model \joe{give fw reference}
, and a \emph{physical timestamp}, recording the wall-clock time of the event. 
Other external events, such as those generated from an automated process in the browser on behalf of the user (e.g., timers), follow the same model.

% A: Why? Does DIEL treat them differently? What happens if I .NewInput to a non EVENT table?
% . \joe{not shown in Fig 2?}
Developers mark event tables with the \dcode{EVENT} keyword (5A).  \sys generates the corresponding JavaScript handler for the API \apicode{NewEvent}.  Developers are free to use their favorite frontend libraries (e.g., Vega-Lite, D3) to generate events; they simply invoke the \sys API to store new events in the event table.

\begin{table}[tb]
\centering
\begin{tabular}{ | l | l | } 
  \hline
keyword   & syntax \texttt{\small CREATE} \\ \hline \hline
\texttt{event} & \texttt{\small EVENT TABLE <table\_name>(<column>)}... \\ 
& \texttt{\small EVENT VIEW <view\_name> AS SELECT}... \\ \hline
\texttt{output} & \texttt{\small OUTPUT <output\_name> AS SELECT}... \\ \hline

% All the tables are managed by \diel at run time.  
\end{tabular}
  \caption{\sys syntax for creating tables.  An \dcode{EVENT TABLE} stores the history of an event.  Developers can access the table with two additional columns maintained by \diel: \emph{timestep} (logic time of the event) and \emph{timestamp} (wall-clock time of the event).  An \dcode{EVENT VIEW} is a named SQL query (view) that spans the local and server databases.  \add{The system appends new view evaluation results to the \dcode{EVENT VIEW}, annotated with the timestep of the corresponding \dcode{EVENT TABLE} entry tracked by \sys, \emph{request\_timestep}, as well as the \emph{timestep} caused by the new event.}   Developers access these views \replace{the same way}{similar to how} they access event tables\del{, with an additional column \emph{request\_timestep}, which indicates when the view was requested}.  An \dcode{OUTPUT} is a view whose results, after each timestep, are evaluated and passed to the rendering function bound via the \apicode{BindOutput} API. \add{These three key constructs, \dcode{OUTPUT}, \dcode{EVENT} and \emph{timestep}s augment SQL with reactive semantics for non-blocking interactive visualizations.}}
\label{tab:relations}
\end{table}

\begin{table}[bt]
\centering
\begin{tabular}{ | l | p{7cm} | } 
  \hline
API    & syntax \\ \hline \hline
input  & \texttt{\small diel.NewEvent('<event>', <object>)} \\ \hline
output & \texttt{\small diel.BindOutput('<output>',   <rendering\_func>)} \\ \hline
  \end{tabular}
  \caption{\sys JavaScript APIs to interface with the frontend. \smcode{NewEvent} takes in events from the developer, e.g., user's selection. \smcode{BindOutput} binds the outputs to a rendering function specified by the developer that takes a table and renders a visualization.
  }
  \label{tab:api}
\end{table}

% \vspace{-2mm}
\noindent\textbf{Interactive Visualizations} User interactions add events to \sys event tables that ultimately update the visualizations. 
% \gray{The data backing interactive visualization updates based on user interactions.}
This happens naturally because the developer can query the \dcode{EVENT} tables in the same way as any normal data table.
Any such queries marked with \dcode{OUTPUT} will automatically generate the corresponding \apicode{BindOutput} API to register a rendering function callback (used in e.g., \replace{5D}{5C}).

When a developer is working with a remote database, they need to query \emph{across} both the local and the remote databases.
For instance, the \smcode{fires} table is on a remote database, whereas \smcode{selection} is in the local database.
\sys automatically manages the asynchronous communication between the two databases. Due to this asynchrony, we treat such queries as events as well, and ask developers to explicitly mark these queries with \dcode{EVENT VIEW} so they are aware of the table being an event log (despite the fact that \sys can automatically detect them). 
For each \dcode{EVENT VIEW}, \sys maintains the corresponding timesteps of the event that caused its reevaluation, and stores the timestep values as a column named \smcode{request\_timestep} in the \dcode{EVENT VIEW}.
Developers then use the data in \dcode{EVENT VIEW}s to specify \dcode{OUTPUT}s, with the help of the timesteps to specify how asynchronous events should be handled in the face of out-of-order events.

For example, the line chart in Fig.~\ref{fig:example} (3C) visualizes the distribution of fires by year.
The query in (5D) fetches the distribution of fires based on the \texttt{\small us\_state} selected: it combines the \smcode{fires} and \smcode{selection} event tables using the \smcode{JOIN} operator, and then filters the rows in \smcode{fires} by applying the filter on the selected \smcode{us\_state} selected using the \replace{\smcode{WHERE} operator}{\smcode{JOIN...ON} operator}.
(5C) is a query that selects the year distribution results of the most recent interaction\del{ is selected,} based on the \smcode{timestep} data.
This ensures that the interface displays the correct data despite out of order responses.

%%%%%%%%%%%%%%%%%%%%%%%

\subsection{\sys  Execution}
\label{sec:exec}

\begin{figure}[tb]
 \centering
 \includegraphics[width=\linewidth]{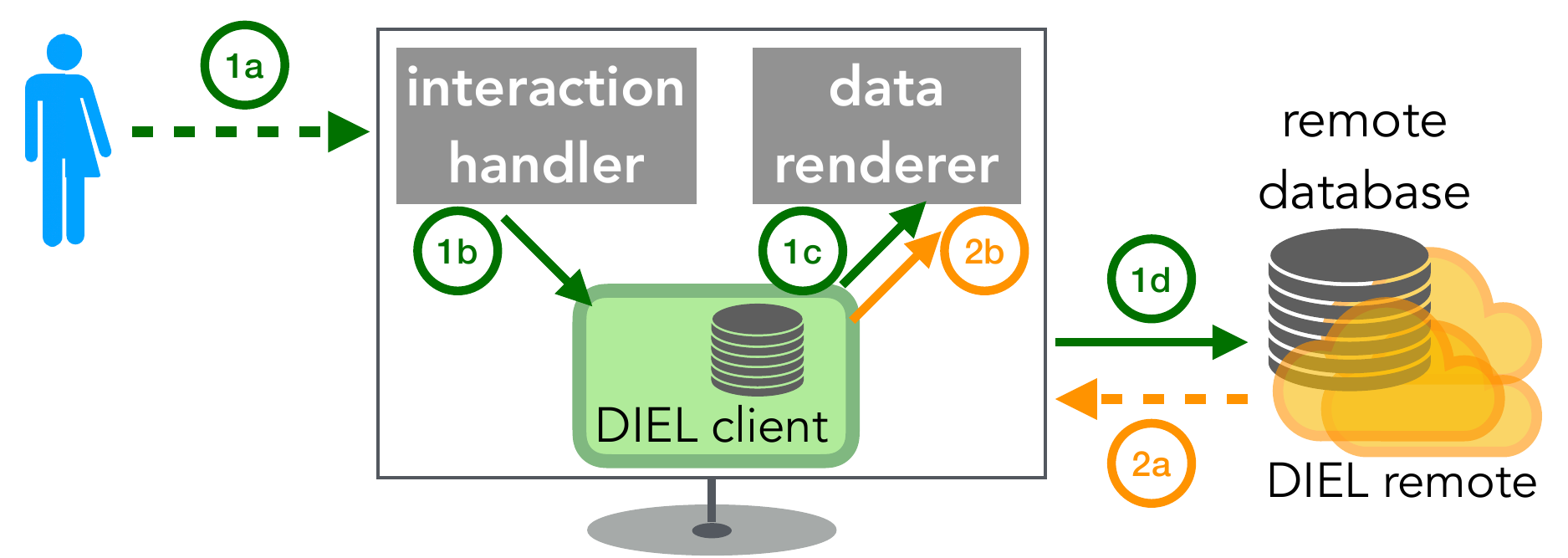}
 \caption{Illustration of the \sys runtime. The dashed arrows indicate asynchronous events that advance the system logical timestep forward. There are two in this diagram, one from user interaction (green) one from a database response (orange).  The solid arrows indicate synchronous evaluation after each new event.}
 \label{fig:runtime}
\end{figure}

% two components, client database and remote database, are involved in event handling
% diel orchestrates these components based on developer specification to handle events
% there are two types of events
% one is interactions
% Another one is data that’s triggered by an interaction event
% In interactions….details here
% In interaction events….details here
\sys orchestrates the local database and one or more remote databases in response to new events.
Each event is evaluated \emph{synchronously} in the local thread, corresponding to one \emph{logical timestep}. As a result,
the application-level effects of events with more recent logical timestamps are guaranteed to occur after events with earlier logical timestamps.
%  more recent timesteps are guaranteed to occur after prior timesteps\ewu{do you mean: the application-level effects of events with more recent logical timestamps are {\it guaranteed} to occur after events with earlier logical timestamps}.
% two events
From the perspective of the local database, user interactions and responses from remote databases are both events, however they are handeled slightly differently, as described in Fig.~\ref{fig:runtime}.
%There are two types of events, one is interactions and another is response from remote database.
%We illustrate the steps in Fig.~\ref{fig:runtime} for the two types of events, respectively.

% \ewu{note: The earlier convention lists the (1a) after the descriptive text.  }
\noindent\textbf{Interaction Event:}  The application creates new events from user interactions (1a), and inserts them into the corresponding \dcode{EVENT TABLE} using the \apicode{NewEvent} API call (1b).   \sys re-computes the local \dcode{OUTPUT} tables, and triggers the corresponding render function callbacks (1c).  Finally, \sys handles any \dcode{EVENT VIEW}s that depend on the new event. If it accesses remote data, \diel first sends the appropriate query to the remote database and handles the response as described below (1d).  Otherwise, \sys executes the local \dcode{EVENT VIEW} query.

% \gray{(1a) A user first interacts with the application, which the developer captures with a frontend library and invokes the \sys API \apicode{NewEvent}. (1b) \sys then insert the event into the corresponding \dcode{EVENT TABLE} in the client database. (1c) \sys reevaluate client-side \dcode{OUTPUT}s and sends their values to the corresponding rendering functions specified by the developer. \sys then identifies \dcode{EVENT VIEW}s that depend on that event.  If the query involves remote data, \diel first sends to the remote database the data that it needs to execute the query, and then request results of the \dcode{EVENT VIEW}s (1d).}

\noindent\textbf{Data Response Event:}
Once the local \sys runtime receives a response from a remote database (2a), which is tagged with the originating event's timestep, it calls \apicode{NewEvent} to insert it into the \dcode{EVENT VIEW}.     Finally, the \dcode{OUTPUT}s are reevaluated, and the results are sent to the application rendering functions.
% \gray{After the \sys database wrapper finishes evaluating the queries in the database, it (2a) shares the results with the client, which invokes the \apicode{NewEvent} function. The client \sys runtime then inserts the values into the client database and (2b) invokes the rendering functions with results from output views.}
% \ewu{WHAT?} is

% The \sys runtime directs query executions on all the databases, both the insertion of relevant data into the databases, and the output of relevant views from the databases.
% , either directly by manipulating the DOM or via libraries such as D3 or Vega(-Lite). 
% If new tasks arrive while the remote database is still processing, the wrapper puts the query evaluation request on a queue.

%! TEX root = ..\diel.tex
\section{A Prototype Implementation of DIEL}
\label{sec:prototype}

% p1 what DIEL takes in and what it outputs
To verify the feasibility of the \sys model, we implemented a prototype.  It ingests a \sys specification and produces a distributed dataflow that is executed in conjunction with the frontend JavaScript libraries and backend databases.
% issues
There are two main challenges in this process.
First, queries over distributed data cannot be executed directly since the computation requires data to be co-located.
Second, the \sys model poses performance challenges.
%  a range of

An implementation should both create a dataflow such that necessary data is exchanged between the local and remote databases, and overcome these performance hurdles to keep the interface responsive.
% We need a mechanism to create a dataflow such that necessary data is exchanged between the client and remote databases. 
% We must overcome these hurdles to keep the interface responsive.
As such, \sys builds the distributed dataflow in four phases. The first three phases address the challenge of distribution, and the last phase address the challenge of performance.
% : \emph{compile} turns \sys spec into an , \emph{catalog} retrieves what tables exist in remote database(s) to inform the distribution data, \emph{exchange} lowers the logical IR to a ``physical'' IR that specifies data exchanges required between local and remote databases, and lastly, \emph{optimize} further transforms the physical IR for improved executions.

% this adds some details about how its done
Two libraries manage these steps: (1) a local (browser-based) library updates the JavaScript state, and (2) a remote library configures access to the remote database.  The developer provides these respective libraries with the connection configurations (e.g., Fig~\ref{fig:example} \circled{4} and \circled{6}).

% compile step
In the first step, \sys parses and compiles the specification into an abstract syntax tree that captures the relational operations between tables. The tree diagrams in Fig.~\ref{fig:example}\circled{2} (left) illustrates that of the running example.
With the tree of operators, \sys can now reason about the data exchange needed to co-locate tables for evaluation.
% intermediate representation (IR) that captures the logical constraints, which does not contain physical data location details, for further manipulation

% catalog
To reason about the exchange, \sys next identifies the tables and their schema in the remote databases, i.e. a table catalog.
% \sys needs to know where the tables that the developers reference live, in order to figure out how to exchange data so that the queries can be executed.
% Catalogs can be obtained, for example, in SQLite, by querying \smcode{sqlite\_master}.
% exchange
Then, with the catalog, \sys identifies queries that involve tables on both the local and remote databases. \sys determines, for such queries, what data need to be exchanged between these databases.  This process transforms the high-level logical specification into a distributed dataflow that can now be executed.

% give an example to make it more concrete
An example of such a transformation done by the current prototype is shown in Fig.~\ref{fig:example}\circled{2} (right). \sys decides to perform the evaluation of \smcode{yearDistributionEvent} on the remote database, thus requiring the \smcode{selection} table to be sent over the network to the remote database from the local database, and the result \smcode{yearDistributionEvent} to be sent over the network from the remote to the local database. The output  \smcode{yearDistribution} is then evaluated within the local database upon each step.
Note that this is not the only distributed dataflow possible.  Alternative implementations of \sys could instrument more advanced algorithms that can further optimize for performance by changing what data needs to be shared over the network. \replace{This can be done by  partitioning}{One example is to partition} the execution of queries into more granular subqueries\add{, and another is to leverage WebWorkers on the local client to parallelize compute}.

% optimization
In the final phase, \sys optimizes the physical execution plan from the previous step. Four mechanisms are used.
% transform the execution plan
%  to avoid wasteful reevaluations resulting from the raw distributed dataflow
% say on the high level why things work
% Fortunately, the \sys model allows us many options for optimization.
% say that it's declarative/constraints
% The \sys spec covers a wide range of executions, leaving a large portion of the computation process to optimize over. 
% Further, the unified language based in SQL can leverage the wealth of database optimization techniques that are available.
% This phase contains more instrumentation and ideas than the previous three phases, and merits an extended discussion.

\noindent\emph{1. Selective output invocation.} Interactive dashboards often contain multiple visualizations, and sometimes ones with many visual elements that are expensive to render.
A naive plan will reevaluate all the queries and re-render all the visual marks, which could block the main browser thread from responding to user events.
To address this issue, the \sys prototype analyzes the execution plan and builds a dependency graph of the \sys queries.  Using the dependency graph, \sys selectively evaluates and invokes rendering functions for only the output views dependent on the current event. 
% \add{The selective update method is instrumented only in the middle-ware layer and does not require dependencies of queries within the connected databases}.

This issue can be addressed via static analysis of~\diel queries at compile time. From the queries, we can extract dependencies across queries, views and tables syntactically. For example, in Fig.~\ref{fig:example}, there are two outputs---\smcode{stateDistribution} and \smcode{yearDistribution}. Looking closely at \ref{fig:example} \circled{2}, note that \smcode{selection} ``flows into'' \smcode{yearDistribution}, but does \emph{not} flow into \smcode{stateDistribution}. This reflects the syntax of those outputs in \ref{fig:example} \circled{5}, where \smcode{selection} appears only in the definition of \smcode{yearDistribution}.  Using this information, \diel knows to \del{only }selectively update \smcode{yearDistribution} when there is a new user interaction (\smcode{selection}).

% Same thing for materialized views. If the data is static, materialized views are
% possible, but if the data is taken from a live database, is DIEL able to manage
% materialized views updated in real-time? This is also related to the issue with
% streaming, is the database changes managed by DIEL somehow, e.g. with triggers?
% This issue is central to the architecture of Tableau as far as I know, and they
% need a special high-performance database to maintain materialized views
% efficiently so it might be harder than described in section 5 unless the database
% is static.

\begin{figure}[tb]
 \centering
 \includegraphics[width=\linewidth]{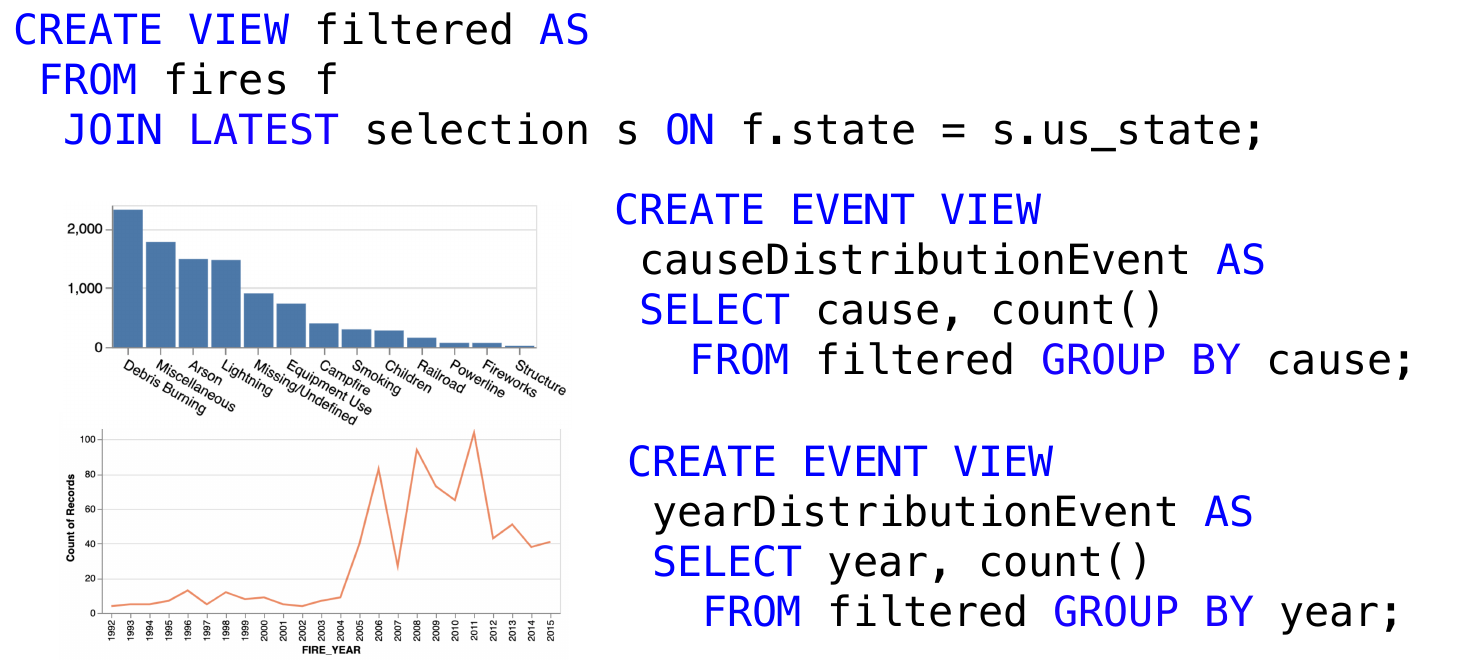}
 \caption{An example of reuse \del{and materialization}: the filtering logic based on the selection in Fig.~\ref{fig:example}, \smcode{filtered}, is shared by two visualizations. \sys analyzes the query plan and materializes \smcode{filtered} to avoid evaluating it twice.}
 \label{fig:reuse}
\end{figure}

\noindent\emph{2. Materialized intermediate views.} 
One interaction can require updating multiple visualizations. Data transformations are sometimes shared when computing these updates.
% Often, one interaction could result in the updates of multiple visualizations.  When computing updates to these visualizations, some data transformations are shared.  
For instance, the \smcode{filtered} view in Fig.~\ref{fig:reuse} is shared by two outputs.
Since views are just named queries in SQL, this can lead to redundant work: whenever the database evaluates any query that references a view, the view is reevaluated as part of the query.  This wasteful reevaluation can be prevented if the view results are materialized into a table for use by downstream queries~\cite{chirkova2012materialized}. For materialized views to work correctly, they must be updated appropriately when the data they query changes. We tackle this problem pragmatically in our prototype: we implement update mechanisms for views in the our local database code; for remote databases, we count on their native support for materialized view maintenance.
% maybe add footnote about who support it

\begin{figure*}[bt]
 \centering
 \includegraphics[width=\textwidth]{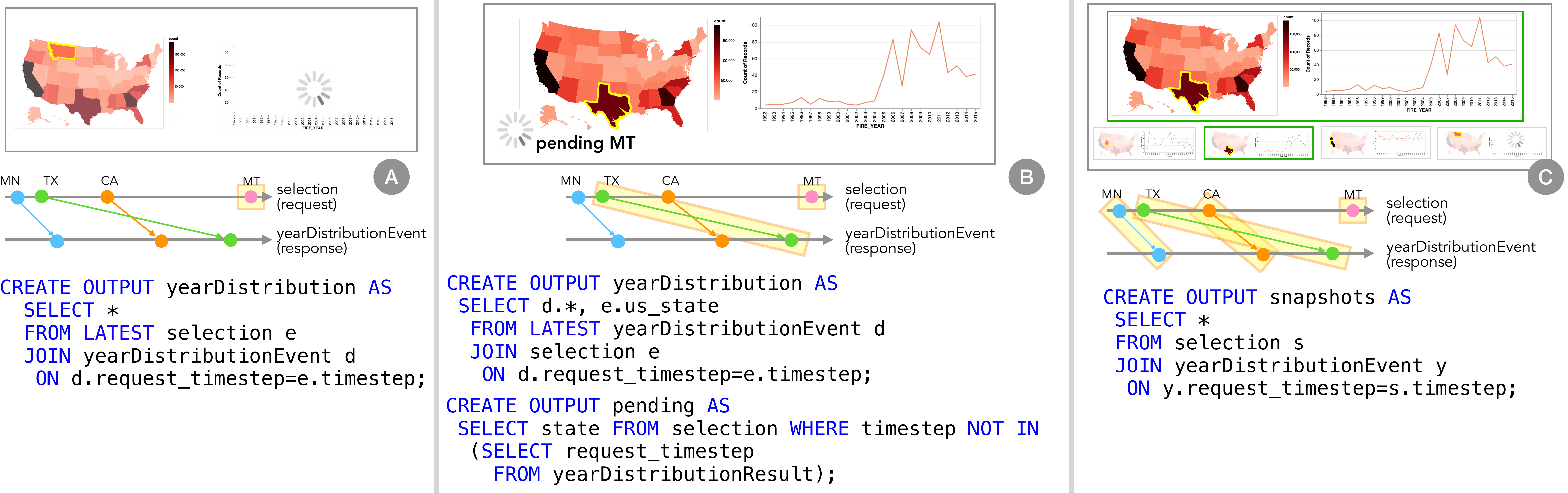}
 \caption{Designs to coordinate asynchronous requests and responses when querying over distributed data:
 \circledgray{A} renders the most recent interaction requested;
%  If the response is not yet received, the \sys \dcode{OUTPUT} \smcode{yearDist} will return zero rows, which the front-end maps to a spinner.
 \circledgray{B} renders the most recent response received as well as any \emph{pending interactions};
%   are recognized separately (and rendered with yellow outlines and a caption)
 \circledgray{C} renders snapshots of all interactions and their corresponding results~\cite{wu2020snapshot}.
%  Rather than selecting the most recent item in the \dcode{EVENT TALBE} \smcode{selection}, the spec instead selects \emph{all} the values.
 }
 \label{fig:async_policy}
\end{figure*}

\noindent\emph{3. Automatic caching.}
Developers often build a client-side cache by hand to amortize the compute and network time to a server.
% To illustrate, consider the sequence of user interactions shown in Fig.~\ref{fig:caching}: when CA is selected by the user for a second time, developers often manually cache the selection results and check for the cache when a new event is made.
For any given specification, \sys can automatically instrument this functionality, thus saving the developers' time.
% how we do it
The server response data are already stored in the event log.  To make use of the past values, \sys analyzes the execution plan at this stage and instruments a layer of indirection to the evaluation of the \dcode{EVENT VIEW}s when a new event arrives:  \sys hashes the parameters of the relevant rows of the event tables and looks up the hash in an \emph{event cache table} instrumented by \sys.
If the hash is present, \sys returns that value, and if not, \sys dispatches the dataflow.
\sys's automatic caching process saves storage by not having multiple copies of the responses, which is especially limited on the client.
In addition to helping save the developers' time, \sys's automatic caching process also alleviates the additional memory usage caused by \sys's log of events.

\noindent\emph{4. Automatic Index Selection.}  
Good database performance is typically tied to building indexes that suit your query workload. For a given \diel program, this workload is known in advance as part of the spec, hence \sys analyzes the query structure to statically determine indexes that will result in good performance.
% one
% \joe{This is redundant with the previous bullet, no? Shorten.}
% not really: we are doing something different
% One technique is \emph{materialized views}, similar to that discussed earlier, but this time applied for the remote databases.  \sys analyzes the overall query structure and programmatically create materialized views in queries executed in remote databases that support it.
% two
% Another technique is \emph{indexes}, which are  auxiliary data structures that speed up data retrieval operations~\cite{ramakrishnan2000database}.  For the client database, \sys analyze the column(s) with predicate clauses and automatically adds an index on the table.  
% We do not add indices for tables tables on the remote database, which would already have indices or be too expensive to build on the fly. 
% (as instrumented by admins) o
% \joe{Do you mean at the client or the server? Either way this is a bit weird ... big tables need indexes even more than small ones no?}

% \begin{figure}[tb]
%  \centering
%  \includegraphics[width=0.7\linewidth]{figs/caching.pdf}
%  \caption{Illustration of \sys's automatic caching.}
%  \label{fig:caching}
% \end{figure}

Besides optimizing the dataflow, we also optimized the execution speed of the local database. Unlike the remote databases, the implementation of the local database is controlled by \diel. The local database runs in the browser's main-thread and any delay would directly block user interactions and HTML updates. Therefore, it is critical that the local database executes quickly.
%   \replace{A recent advance in browser technology, WebAssembly, allows for fast execution of SQLite, which we use as the local database, through optimized transpilation.}{
We make use of a recent advance in browser technology, WebAssembly~\cite{webassembly}\del{, for our local database}. It provides fast execution of SQLite  in the browser by compiling C code to WebAssembly.

% \footnote{\add{The Emscripten~\cite{emscripten} library compiles the SQLite C compilation to WebAssembly, which runs in the browser while preserving C program's fast performance}}.

The above highlights the most novel aspects of the prototype implementation.  Other details, such as the full \diel syntax specification, along with additional language features discussed in Section~\ref{sec:cog}, are detailed in supplementary materials.

\section{Evaluating \sys's Expressivity}
\label{sec:examples}

% \jmh{Maybe the way to address the caveats in the next paragraph is to establish us up front the
% goals for evaluating expressivity of DIEL. Something like ``The role of \sys is to serve as middleware for asynchronous interactions between the user's choice of a static front-end visualization library (e.g., Vega) and a backend data source---in our prototype, a relational database system. To assess the benefits of \sys, we focus on a variety of challenges that arise in managing asynchrony and distribution, including (category 1, category 2, ...). Note that these challenges are independent of the choice of visualization per se; as a result our examples maintain the visualizations from the running example of Fig.~\ref{fig:example}. Etc.....''}

% To assess the benefits of \sys, we show example interactive visualizations that deal with a variety of challenges that arise in managing asynchrony and distribution, including coordinating responses from remote servers, streaming data, composing interactions, and visualizations of interaction history.

To assess the benefits of \sys, we show examples of interactive visualizations that deal with the variety of challenges that arise when managing asynchrony and distribution, including coordinating responses from remote servers, streaming data, and composing interactions.
% , and visualizations of interaction history
% how we are going about this
Our discussion focuses on the aspects addressed by \sys and omits implementation details of the frontend.
% which serves as middleware for asynchronous interactions between a backend database and the user's choice of a front-end visualization library.
We also show how \sys specifications compose in a modular manner, by building each new interaction on the running example of Fig.~\ref{fig:example}. 
Since the challenges are independent of the particular choice of visual encodings, we do not focus on varying the visual designs.
% across examples 

\begin{figure}[tb]
 \centering
 \includegraphics[width=\linewidth]{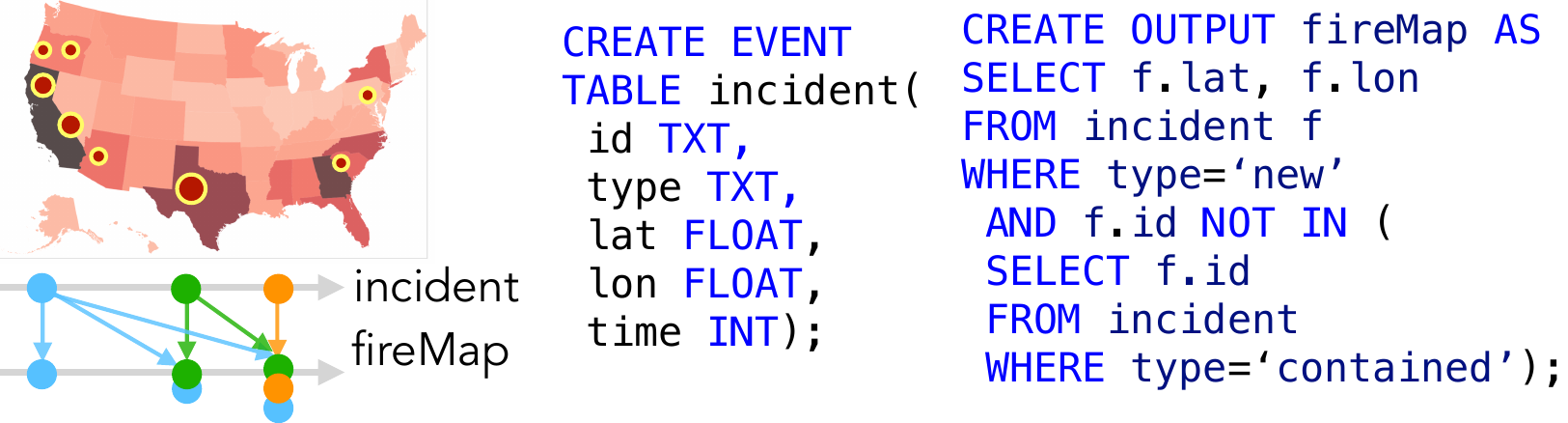}
 \caption{Example \sys spec for the symbol overlay of active fires, determined by selecting \smcode{incidents} that do not yet have a row with the column \smcode{type} of \smcode{contained}.}
 \label{fig:streaming}
\end{figure}

% to support streaming.  Through \sys spec, the runtime derives the output views dependent on the streaming event, both present and past, and calls the corresponding rendering functions automatically

\noindent\textbf{Coordinating Requests and Responses.}  Latency from remote databases' responses can cause inconsistencies in the interface, if not handled properly~\cite{wu2016devil}.
For instance, Fig.~\ref{fig:async_policy} shows a timeline of user interactions (\smcode{selection}) and server responses (\smcode{yearDistributionEvent}).
The user clicks \smcode{TX} and then \smcode{CA}, but the remote database responds with the result for \smcode{CA} first. Rendering the most recent result (\smcode{TX}) would surprise a user who is expecting \smcode{CA}.
% The user clicks \smcode{TX} and then \smcode{CA}, but as it happens, the remote database responds with result for \smcode{CA} before \smcode{TX}. Rendering the most recent result (\smcode{TX}) would be surprising for the user (who is expecting \smcode{CA}).
To avoid inconsistent interfaces, developers can use \sys to specify a range of designs in a few lines of code.  We walk through three possible designs shown in Fig.~\ref{fig:async_policy}.

Option \circledgray{A} always displays the result of the most recent interaction.  The \sys spec first identifies the interaction by selecting the rows with the highest timestep (\dcode{LATEST} \smcode{selection}), then selects the rows in \smcode{yearDistributionEvent} with matching request-response timesteps (\smcode{d.request\_timestep = e.timestep}).  If no match is available yet, nothing is returned.  The front-end visualization logic could indicate as such, e.g., using a spinner as shown.

Option \circledgray{B} always displays the most recent response and its corresponding selection, as well as pending selections.
The \sys spec first selects the most recent response (\dcode{LATEST} \smcode{yearDistributionEvent}), then joins it with the \smcode{selection} table to retrieve the corresponding value of the selection (\smcode{e.us\_state}) by matching their timesteps.
The second output \texttt{pending} represents pending selections and is computed by finding the request(s) that do not have a corresponding response.
% space saving
% In this particular sequence of events, \texttt{TX} is the selection with the most recent response, and \texttt{MT} is the pending selection.

%  request/response latencies developed in recent research, \emph{interaction snapshots} 
Option \circledgray{C} displays ``snapshots'' of all interactions~\cite{wu2020snapshot}, where past interactions and their results are scaled down into a scrolling pane of small multiples at the bottom.  The \sys spec simply selects all the interactions from the event table \smcode{selection}, joined with the corresponding responses by their timesteps. The snapshots allow a user to interact with the visualization and navigate to prior states, concurrent with the loading of new responses.

% A Can you say more about what they would need to store/annotate specifically. And what makes that difficult/error-prone. i.e., do more than state these things to be true, but rather say _why_ they are true.
Without \sys, implementing these designs would require the developer to manually keep track of events\,---\,store the points of interest selected, their respective responses, and the global ordering of all the events\,---\,and coordinate multiple event handlers.  While each step is simple in isolation, put together the complexity of this low-level data-recording and event handling compounds substantially.  Developers may have trouble reasoning about the \emph{overall} design.

% comparison
% With \sys, the experience is different. 
\sys, on the other hand,
% A *briefly* say how DIEL enables things. You might find this repetitive with what you've said in prev sections but, remember, this is all new material for your reader so they'll need things to be stated a few times in order to remember.
% A You can even point to specific snippets of the DIEL spec (e.g., matching d.request_timestep to e.timestep) as your example for %two.
% For one, \sys, 
encourages a consistent experience by asking the developer to specify which of the events should be in the output at any given time. There is no accidental design resulting from interrupt-driven event handling, such as immediately rendering whatever response arrives.
% the user experience is not mere side effect of low-level event handling as an afterthought
% two
Furthermore, \sys takes care of recording event history and provenance. The developer can query these data directly\,---\,for example identifying the responses from remote databases with timestep data that \sys automatically maintains.
% \smcode{d.request\_timestep} with \smcode{e.timestep}.
As a result, the developer can iterate on alternate designs without \del{any }instrumentation overhead.

% The examples we explored are all a few lines of \sys specification, which is in contrast to traditional imperative event handling where some, perhaps more user-friendly, designs are avoided due to their additional complexity.

\begin{figure}[tb]
 \centering
 \includegraphics[width=\linewidth]{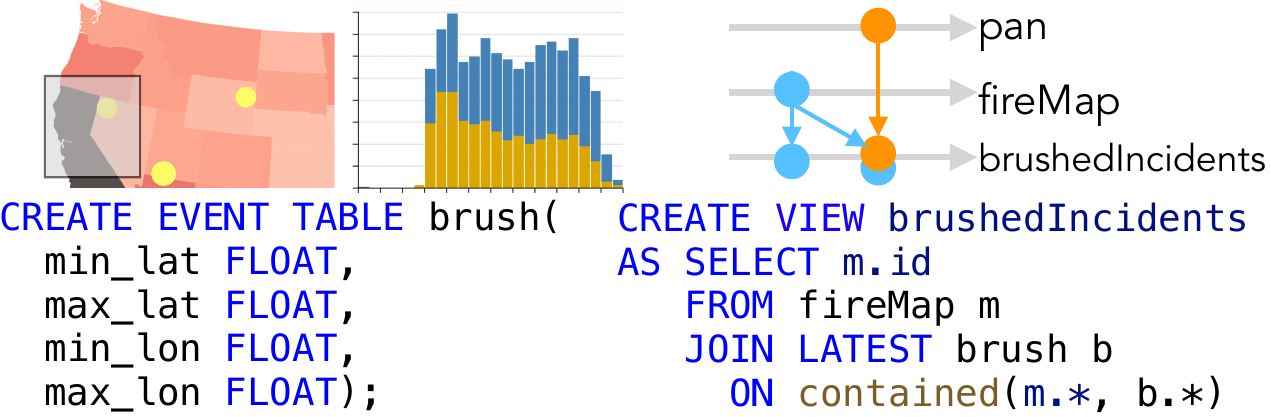}
 \caption{Example \sys spec for brushing interaction: \smcode{brushedIncidents} selects fires in the symbol map \smcode{fireMap} that fall\del{s} into the brushed region \smcode{pan} to update the bar chart (query omitted). \add{The user-defined function \smcode{contained} helps make the code more concise.}}
 \label{fig:compose_brush_stream}
\end{figure}
%%%%%%%%%%%
\noindent\textbf{Streaming Data.} 
%  Important to get visualization terms correct throughout. The base layer in a choropleth (not a heatmap). The overlay is a symbol map (not a scatterplot).
% Working with streaming data may require the developer to  to perform further computation over.
Given the real-time nature of fires, the developer may want to incorporate streaming data into their visualization.  Fig.~\ref{fig:streaming} overlays the choropleth in Fig.~\ref{fig:example} with a symbol map of active current fires.
The event \smcode{incident} contains the location of the fire and whether it is new or contained.
When a new \smcode{incident} event arrives, fires are either added to or removed from the symbol map overlay.
% Consider my writing advice from earlier/B2: avoid sentences that make statements like "this is easy" because they're likely to annoy your reviewer/reader without additional qualification/evaluation (easier for whom? how are you sure that it's making it easy?).
% Rather, statements like this should not be necessary because the rest of the writing/example should _demonstrate_ that the DIEL spec makes it easy. i.e., show rather than tell.
% And so, if you drop this sentence, consider it an additional prompt/pressure to make sure your writing is really richly conveying how DIEL does indeed make it easier.
% The developer can  raw stream data and output the result to the UI.  
To implement this design with \sys, the event \smcode{incident} is captured as an event table and the data for the overlay is captured by the output table \smcode{fireMap}.  Each tuple from the latter is used to query the \smcode{incident} table to identify fires that have not yet been controlled, via the \smcode{NOT IN} subquery.
Note how the developer can rely on a few lines of \sys code, instead of programming custom JavaScript functions to store and manipulate streaming events\footnote{As with materialized views in Section~\ref{sec:prototype}, we take a pragmatic approach in our prototype: we handle streams in the client database, but do not support streaming in the remote databases.}.

\noindent\textbf{Composing Events: Interaction and Streaming.}
Cross-linking is a common interaction technique~\cite{shneiderman2003eyes}. Fig.~\ref{fig:compose_brush_stream} shows a new brush selection added to the map visualization, linked to the bar chart \del{shown to the right }(the backing query is not shown for brevity).
\smcode{contained} checks if the \smcode{lat, lon} \replace{bounds}{values} in \smcode{fireMap} are within the \smcode{min} and \smcode{max} \add{bounds} of the brush. It is a utility function defined by \sys, using the \emph{user-defined function} (UDF) construct in SQL~\cite{ramakrishnan2000database}. \add{UDFs are supported by the frontend database library we use, \smcode{sql.js}~\cite{sqljsUdf}, and developers can define UDFs through the \sys runtime API, \smcode{AddUDF}}.

The new interaction composes with the streaming \smcode{firemap} view from before\,---\,if there is a new event received that falls into the brushed area, the incidents selected in \smcode{brushedIncidents} will also be updated, and any dependent output views will be updated as well.  
This subtle instrumentation, automatically performed by \sys, may be difficult for a developer to catch in a traditional implementation where the logic may be dispersed into different event handlers.

\begin{figure}[tb]
 \centering
 \includegraphics[width=\linewidth]{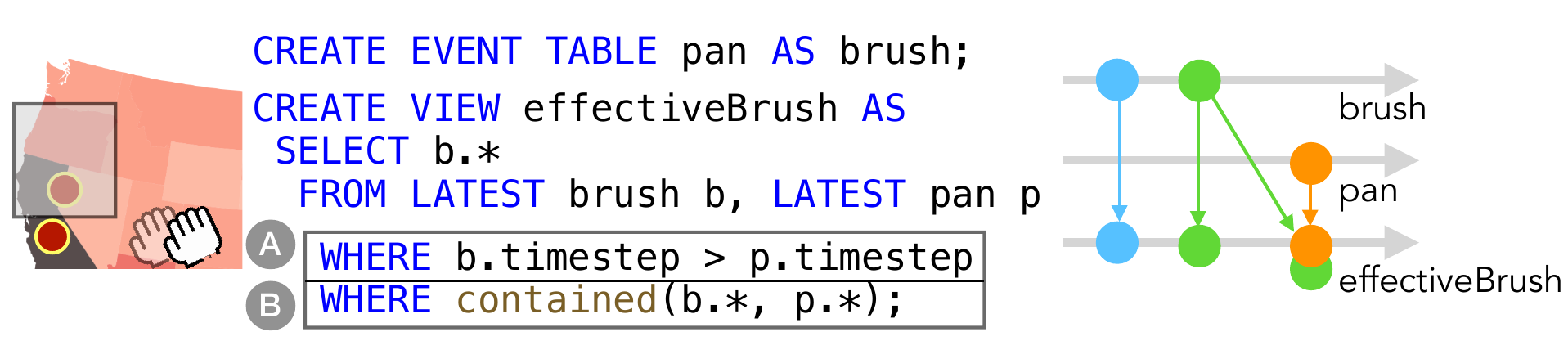}
 \caption{Example \sys specification of composing two interactions: panning and brushing (from Fig.~\ref{fig:compose_brush_stream}).  There are two ways to coordinate: \circledgray{A}  removes the brush selection whenever there is a more recent panning event, and \circledgray{B} removes the brush only when \replace{the brush}{it} is panned out of view.}.
 \label{fig:pan_brush}
\end{figure}

%%%%%%%%%%%
\noindent\textbf{Composing Events: Brushing and Panning.}
Different interactions serve different purposes and more than one interaction could be employed for the same visualization, which the developer may need to coordinate.
Following the running example, suppose now the developer wishes to introduce a pan-zoom interaction, shown in Fig.~\ref{fig:pan_brush}. The \smcode{brush} table is defined in geographic coordinates, so 
the user can pan the map to an area where the brush is no longer visible, and the value of the selection is in question.
We present two possible designs to address this ambiguity. Both derives a new brush, \smcode{effectiveBrush} for use in place of the raw \smcode{brush}.

% one
Option \circledgray{A} invalidates the brush when the user initiates a new panning interaction.
The \sys spec selects the most recent brush (\dcode{LATEST} \smcode{brush}) only if it occurred \emph{after} the most recent pan (\dcode{LATEST} \smcode{pan}).
% two
If we replace option \circledgray{A} with
option \circledgray{B}, we instead invalidate the brush only when a new panning event moves the brush \emph{out of view}.
In either case, developers do not manually modify the callbacks to the panning interaction handler to check and remove the previous brush selection. The ``removal'' is specified declaratively and enforced implicitly by \sys as the logical timesteps progress.

\begin{figure}[tb]
 \centering
 \includegraphics[width=\linewidth]{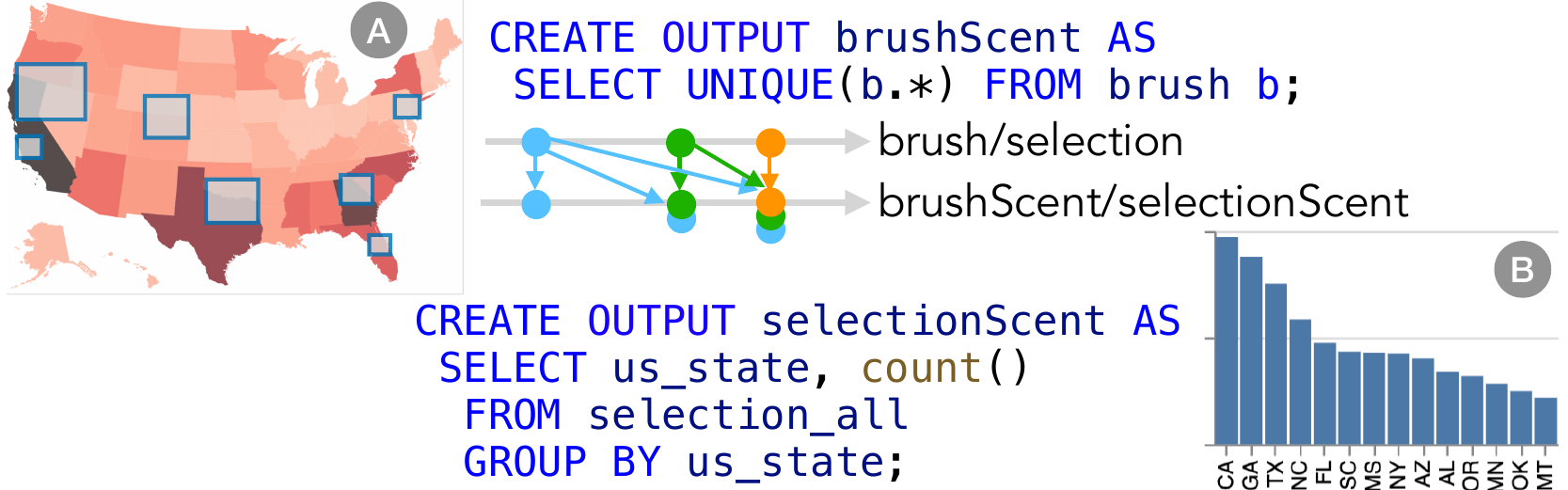}
 \caption{Example \sys spec for \circledgray{A} Hindsight: select all unique brushes in the \smcode{brush} event table, and \circledgray{B} Scented Widget: a bar chart that shows the frequency of selections across users.}
 \label{fig:scent}
\end{figure}

%%%%%%%%%%%
\noindent\textbf{Interaction Scent.} Research has shown the benefit of visualizing interaction history \add{to provide a visible trail (``scent'') of prior interactions.}
%  could help augment the user's cognition by showing past interactions.
\emph{HindSight} visualizes the user's prior interactions in the visualization~\cite{feng2017hindsight};
\emph{Scented Widgets} visualize interactions by other users~\cite{willett2007scented};
\emph{Interaction Snapshots} visualize historical interactions and their results~\cite{wu2020snapshot}.
% example
We have shown an example of \emph{Interaction Snapshots} in Fig.~\ref{fig:async_policy}; we now discuss the other two designs in Fig.~\ref{fig:scent}.

\circledgray{A} instantiates a \emph{HindSight} design, where all prior brushes are shown~\cite{feng2017hindsight}.
The \sys spec for \smcode{brushScent} selects unique brushes through the \smcode{UNIQUE} operator.
\circledgray{B} shows an example design of \emph{scented widgets}, where a histogram of all prior \smcode{us\_state} selections are aggregated and counted~\cite{willett2007scented}.
The \smcode{selection\_all} is a table in the database that stores all the event tables from each session,  which is saved by the developer from event tables to the remote databases (and not handled automatically by the current \diel prototype), e.g., \smcode{INSERT INTO selection\_all SELECT * FROM brush}.
% \add{An alternative scent design \circledgray{B} is similarly implemented through basic relational operator, \smcode{GROUP BY}, whereby the widget presents the frequency of selections made prior.}

With \sys, the data backing these visualizations is already in event tables ready to be queried, and the visualization scents are automatically updated.  Without \sys, the developer would have to manually store the events and re-render the visualization scents after each interaction, or use libraries like \emph{Trrack}~\cite{cutler2020trrack}.
% These history-based designs require additional instrumentation with existing visualizations, such as . With \sys, interaction history is built in.  
%  used to annotate the map visualization

\section{Evaluating \sys's Performance}
\label{sec:perf}

% i.e., what did you run; what did you measure; what were the results; why were the results what they were

We evaluate the feasibility of the \sys model using our prototype implementation. 
% \replace{Given that \sys introduces an additional middleware layer written in a high-level language, we are concerned about its impact on initial setup and on its effects on the interactivity of event handling. Given that \sys offers the ability to use a full-service database running outside the browser, we also evaluate whether \sys does indeed scale as the dataset grows in size.}{
We focus on the overhead that the \diel middleware introduces during initial setup and user interactions, as well as its ability to scale to large datasets as compared to leading visualization libraries.

% measure three aspects of performance: the time taken for initial setup; the time taken for event handling; and how \sys scales as the dataset grows in size.
% to handle events in a quick and scalable manner.
% We first decompose the time \sys takes at the initial setup.
% We then decompose the time when \sys handles each event.
% We then show how \sys scales as the dataset grows in size.
% the performance breakdown for both the initial loading delay and the performance for each time step.  We then contrast \sys's ability to scale to larger datasets to that of Vega's by comparing the processing ability and time for increasing sizes of data over an equivalent interactive visualization.
% details
We used Kaggle's wildfire dataset~\cite{kagglefire}, creating the visualization shown in Fig.~\ref{fig:example}, and conducted evaluations on a MacBook Pro with 2.7 GHz Quad-Core Intel Core i7 and 16 GB of memory.  In the experiments, the ``remote'' database server is a SQLite process running on a web application deployed on Heroku ("Free Dynos" tier with 512 MB of RAM). The network bandwidth was 18.5 Mbps.
Changing the dataset size or database will not affect the overhead that the \diel middle-ware incurs. Thus we chose to focus on \replace{these three}{the above} evaluation configurations.

% \footnote{
% \replace{
% An evaluation against more performance databases and larger datasets is plausible but does not add more information to the evaluation, since the differences would solely be on the databases and independent of the \sys middle-ware.
% .}{
% Changing the dataset size or database simple changes to fixed costs that any implementation would incur, and our findings that the DIEL middle-ware incurs minimal overhead would not change.
% }

% \footnote{The code for the experiments with two alternative implementations are included in supplementary materials}.

\begin{figure}[tb]
  \centering
  \includegraphics[width=\linewidth]{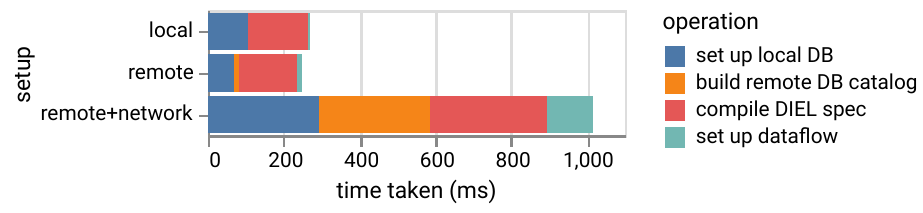}
  \caption{The x-axis represents the respective time taken for \sys's initialization steps, in milliseconds, and the y-axis represents the type of setup being measured.}
  \label{fig:perf_setup}
\end{figure}

\noindent\textbf{Initialization.} Fig.~\ref{fig:perf_setup} shows the time taken to setup \sys.
% We now unpack the two cases.
The \emph{local} setup only involves a local database in the main browser thread.
\sys (1) 
% \setulcolor{steelblue}\ul{
sets up the synchronous database in the main thread; (2)
% \setulcolor{crimson}\ul{
compiles the \textsc{diel} spec into logical representations, including optimization steps such as caching; and, (3) 
% \setulcolor{emerald}\ul{
sets up the views in the local database.
% As the local database contains all the data to be computed, step (1) takes time.  It would take longer if the data loaded is larger.
The \emph{remote} setup accesses the remote database server.
\sys shares step (1) and (2) of the local setup, then it 
% \setulcolor{carrot}  \ul{
builds a \emph{catalog} of tables in remote databases and 
% \setulcolor{emerald}\ul{
sets up a distributed dataflow based on the catalog.
% the first three steps as before, but performs step (4) differently. It .
%  in the remote database
% Step (4) takes extra time to build the catalog and set up the distributed dataflow.
% \replace{The \emph{remote+network} case is the same as remote but involves  sending data over the network (as opposed to running on the same machine). As expected, this introduces additional latency due to networking. For instance, in ``set up local DB'', the latency is 200 ms higher than remote because of the time it takes to send the SQL.js javascript library to the browser page. The remote DB catalog also requires a websocket message round tripped over the network. The latencies would be shorter over faster bandwidth and vice versa.}{
The \emph{remote+network} results are the same as \emph{remote}, but with the addition of nontrivial network communication overheads.  These introduce fixed costs---such as the 200ms overhead to fetch the SQL.js library in order to set up the local database---as well as variable costs to send results between the client and server.  These costs depend on the network bandwidth. 
For instance, decreasing the network bandwidth to 1.4 Mbps caused the set up local database to be 8687ms.

In all three cases, we see that the initialization time does not pose a usability challenge, especially given that the initiation is a one time cost at the beginning of loading the web page containing the \del{interactive }visualizations.

% It would take longer if the network is slow.
% Of course, some steps could take longer: local database setup, if a large amount of data is loaded into the local database; building a catalog of remote databases, if the network is slow; and compilation, if the \sys spec is complex. 

\begin{figure}[tb]
  \centering
  \includegraphics[width=\linewidth]{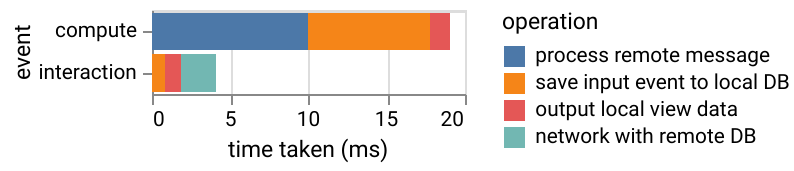}
  \caption{The x-axis represents the time taken for the event handling steps, and the y-axis the type of event being measured. These results are evaluated against both the \emph{remote+network} and the \emph{remote} only setup. The numbers follow the same distribution because the tick performance is decoupled from the database query processing performance.  The steps are all \emph{local} evaluations after \add{receiving} the asynchronous events \del{have been received}. The actual latencies for each individual response are depicted in Fig.~\ref{fig:perf}.}
  \label{fig:perf_tick}
\end{figure}

% You need to describe what the two events are, and what the benchmark test actually does/measures, similar to what you did for initialization. i.e., what specific steps were executed?
\noindent\textbf{Event Handling.}  Fig.~\ref{fig:perf_tick} shows the time taken to handle events.
% details
One type of event is user interaction, which took less than 5 ms to handle.  During that time, \sys (1) 
% \setulcolor{carrot}\ul{
saves the input event to the local database (serializing from JavaScript to SQLite); (2) 
% \setulcolor{crimson}\ul{
outputs the view data (deserializing from SQLite to JavaScript); and, (3) 
% \setulcolor{emerald}\ul{
networks with the remote database.
% two
Another type of event is a remote database response, which takes about 20 ms to handle.  During that time, \sys (1) 
% \setulcolor{steelblue}\ul{
processes the remote message; (2) 
% \setulcolor{carrot}\ul{
saves the input event to the local database; and, (3) 
% \setulcolor{crimson}\ul{
outputs the view data.
Processing the data from the remote database and serializing that data into the local database takes the bulk of the time.
% The time would be a little longer if the workload is large in volume. 
We see that the overhead in either case is well under the limit of 100 msec prescribed by Card et al. to sustain perceptual causality~\cite{card1986model}.

\noindent\textbf{Different Data Sizes.}  To demonstrate a \sys program's ability to scale, we created samples from 10 thousand rows to 4 million rows and benchmarked two implementations of the visualization shown in Fig.~\ref{fig:example}: one with \sys (the SQLite server and static Vega visualizations) and one with Reactive Vega running in the browser.
% , with the same visualization specs.
We measured the time taken between the handler receiving the interaction and outputting the computed result.  This excludes the rendering logic, which is common across the experimental runs.  The SQLite and \sys results also exclude the network latency since it is shared in both cases.

\begin{figure}[bt]
 \centering
 \includegraphics[width=0.7\linewidth]{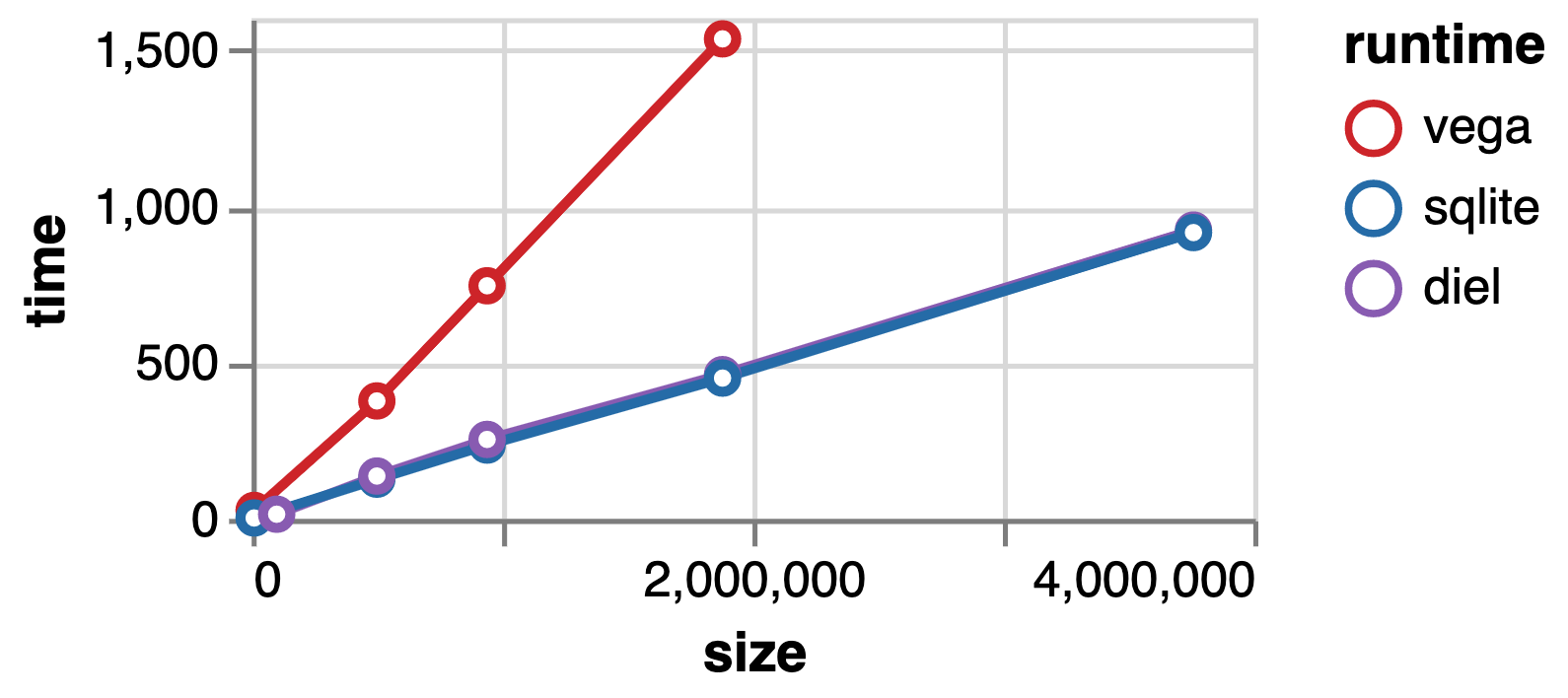}
 \caption{A visualization of how long the query processing takes as the data grows in size.  
 One program is implemented with Vega, another with \sys used in conjunction with a \emph{remote} SQLite database, whose raw query time is also plotted.
%  Comparing the implementation in \sys with that in Vega, w
 \sys is able to make use of the backend database and process data beyond the capacity of the browser, with less than 1s of additional overhead for (de)serialization when working with the remote database.}
 \label{fig:perf}
\end{figure}

% \replace{The result, shown in Fig.~\ref{fig:perf}, is as we had hoped. We see}{
Fig.~\ref{fig:perf} shows that \sys is able to handle the increasingly large data by leveraging resources beyond the client, whereas a client-only Reactive Vega application freezes the browser at 4 million rows of data.
% another
We also measured the query evaluation time directly against the remote SQLite database\,---\,\sys's processing time follows closely with negligible serialization/serialization overhead.
Additionally, the \sys implementation benefits from a \emph{non-blocking} design because it does not take up costly resources on the main thread.

% A It is really bizarre to say an evaluation is incomplete. That's an immediate red flag. I think you mean to say that we're not exhaustively analyzing DIEL's performance. And then you want to defend that decision: the core motivation  ,
While this study does not exhaustively analyze \sys's performance, we believe it addresses \sys's core motivations of usability, expressivity and scalability via remote data servers.
% Note that the colocation of the remote database server on the same machine as the browser was a stress test of sorts, focusing our attention more directly on \sys's latency. 
% Had we placed the remote database on a distant server (say in the cloud) or scaled to a massive database, the main change would have been increased latencies for responding to asynchronous database requests\,---\,i.e. further motivation for the asynchronous functionality provided by \sys.

% .  We could cover other factors like network bandwidth, databases used, and the queries being executed, 

% The graph is proof of the prototype's ability to scale using existing back-ends.  We chose the least sophisticated database---SQLite, which, compared to other databases, is not optimized for analytics.  If we use \sys with a faster database, such as Postgres or a GPU powered database, the performance will improve and follow that of the database used.
% Note that the performance of the prototype largely depends on which database is currently executing the query.  For instance, a SQLite instance versus a GPU powered database will execute the same query differently.
% There are limits however: as a general-purpose model, \sys will not be able to compete with custom optimized systems, e.g., Falcon for cross-filtering~\cite{moritz2019falcon}. Though it is interesting future work to scaffold progressive visualization techniques into \sys.

\section{\replace{Evaluating}{A Discussion of} \sys's Usability}
\label{sec:cog}

We evaluate the programming experience of \sys using the \emph{Cognitive Dimensions of Notation}~\cite{blackwell2001cognitive}, a set of considerations to evaluate the effectiveness of notational systems.  Cognitive dimensions have been used by prior visualization systems to evaluate their usability ~\cite{satyanarayan2014declarative, satyanarayan2014lyra, zong2020lyra, satyanarayan2019critical}.  We pick a relevant subset and contrast the effectiveness of \sys against that of current common practices.

%%%%%%%%%%%%%%%%%%%%%%%%%%%%%
\noindent\emph{\textbf{Viscosity} (resistance to change)}.  \sys follows relational database abstractions and benefits from its ease \replace{for}{of} change. First, its declarative query specification abstracts \emph{what} to compute from \emph{how} to compute \add{(known as ``physical data independence'' in database literature)}. As a result, switching from a client-only application to a client-server one (or to one using WebWorkers) requires only a few lines of change to configuration details.
Fig.~\ref{fig:example}(4E) provides an example of the configuration, \smcode{remoteDbConfigs}, a JavaScript object containing high-level specification such as the type of the database used and what socket to connect to the \sys database wrapper.
Query changes are moreover isolated from table changes \replace{thanks}{due} to the \replace{classical}{``logical} data independence\add{''} property of the relational model~\cite{ramakrishnan2000database}.  For instance, if the \smcode{incident} schema changes because the data provider now also includes the reporting station, none of the downstream queries\del{, \smcode{fireMap} or \smcode{pannedIncidents},} would need to change\del{; more complex schema changes can be hidden behind relational views}. 
Finally, views also provide a way to reuse logic within an application.  
Fig.~\ref{fig:reuse} shows an example where a change to the internal specification of the query \smcode{filtered} would be abstracted away from the dependent views as long as its \smcode{SELECT} clause is unchanged.
\add{These properties combine to make more complex interactive visualizations easier to author, since multiple components could be reused and changed easily. Consider the running example of visualizations over the wildfire data: they could be combined easily into one visualization dashboard.}

%%%%%%%%%%%%%%%%%%%%%%%%%%%%%
\noindent\emph{\textbf{Closeness of Mapping} (closeness of representation to domain)}.  Since SQL is widely used in modern databases and for general data manipulation tasks~\cite{sosurvey}, \sys closely represents the data processing domain.
However, \sys does not fully represent the complex domain of data visualizations.
For example, the Vega authors identify that visualizations involving small multiples often require hierarchical structures with second-order quantification~\cite{satyanarayan2016reactive}.

We agree that the expressiveness of SQL\add{, and hence \sys,} is more limited than interactive visualization frameworks\add{, which are implemented in turing complete languages}.  
Relational languages like \sys express\del{es} only first-order logic.
% \footnote{The ability to define and quantify relationships between entities (such as expressing the query ``given a set of fires, find all fires in NY'')}.
In particular, \sys can define and quantify relationships between entities, but cannot quantify over a data-dependent set of table names or column names.
% Can we provide an example that is relevant to DIEL use cases?
% \add{This limits the expressiveness of \sys}.
Yet this limitation is key for distributed execution and optimization.  Neither physical data independence nor the rich optimization methods discussed in Section~\ref{sec:rational} would be easy to implement without the relational abstraction.
\add{In this sense, any alternative approach towards distributed execution at scale will end up using the same principles and hence facing similar limitations.}
% , so it is not a limitation that DIEL needs, but one that scaling out in general leverages
% A You'll need to connect this to DIEL more closely... i.e., if someone wants to do custom data manip, if they're forced to do it in the client, doesn't that return them back to the problem state setup in the paper's introduction?
% Furthermore, this does not need to be a dichotomy.
Moreover, developers could transform the data into other forms by manipulating the output tables provided by \sys in any JavaScript functions.  For instance, they can use Vega to transform the tabular data into nested groups to render small multiples.
These data transformations happen at the end of \sys's dataflow and does not diminish the effectiveness of \sys's ability to orchestrate \del{distributed }query evaluations \add{across the client and remote databases}.
% support the application, which is primarily around distributing data between the client and remotes and coordinating events.
% However \sys does not \emph{constrain} the developers because they are expected to use \sys as a library.  They can make other types of data manipulation on the client, either through more .
% A: Note, it's fine (even good!) to be only critical along some of these dimensions and doesn't weaken your contribution at all. So don't feel pressured to have a defense for every critique you can levy. If it is a fair critique/analysis, then the writing should accept it and justify why the tradeoff is worth it, and what alternatives there are (this sentence is getting to the latter, but it's a little defensive right now).

%%%%%%%%%%%%%%%%%%%%%%%%%%%%%
% \red{NEED WORK}
\noindent\emph{\textbf{Consistency} (similar semantics are expressed in similar syntactic forms)}.  In \sys, the only data structure is a table.  
One of the key contributions of \sys's model is
to unify both ``live'' events and ``stored'' data in 
a single frame of reference---tables---which store both data and history. Once events are reified as data in an event table, \sys presents a unified data-centric language.
% \joe{}

% \gray{We tease out the data aspect of interactions (the points of interest) and the visualization state to unify the representation of interactions, visualizations, and data.  A consistent expression over all the entities involved gives \sys an unique advantage to simplify the programming model and implement end-to-end optimizations.}
% Not only is \sys internally consistent, but it also makes it easier to interface with frontend libraries such as Vega, whose APIs provide declarative su.
% allows for .Having a consistent 
% While the \sys specification would need to be integrated with the logic for pixel space in JavaScript, which is a separate space of reasoning, the contract between data and visual components is well established, as seen in the grammar of graphics and other visual programming systems~\cite{olsten1998viqing, livny1997devise}.
%  When specifying the UI state, both events and the underlying data are queried using the same operators.
% Contrast this model with the traditional approach where developers have to work with different data structures with custom properties.

%%%%%%%%%%%%%%%%%%%%%%%%%%%%%
\noindent\emph{\textbf{Premature Commitment} (constraints on the order of doing things)}.
For \sys to be effective, it imposes a premature commitment.
Developers must represent the state of visualizations using tables
(rather than arbitrary data structures) upfront. 
This premature commitment can hamper a rapid prototyping process, but we believe the advantages of the table format outweigh these concerns.
As discussed in Section~\ref{sec:rational}, the table format facilitates working with distributed data, and makes explicit possibly concurrent processes.

% J COULD BE crisper
%%%%%%%%%%%%%%%%%%%%%%%%%%%%%%
\noindent\emph{\textbf{Role-expressiveness} (the purpose of an entity is readily inferred)}. \sys reuses an existing, well-established data model of relational tables and queries, and introduces only two additional constructs:  \dcode{EVENT}s and \dcode{OUTPUT}s.
This approach has proven sufficiently expressive, as demonstrated by the examples in Section~\ref{sec:examples}. \sys operates as a middleware layer between the frontend and backends, and lifts the logic of data exchange between client and remotes.
The relatively small surface area of \sys's abstractions stands in contrast to the existing imperative code written to support such use cases, which is often dispersed across custom functions with a commensurate burden of role-expressiveness.

%%%%%%%%%%%%%%%%%%%%%%%%%%%%%
\noindent\emph{\textbf{Hidden Dependencies} (important links between entities are not visible)}.
\sys makes dependencies quite explicit. The only type of dependency \sys introduces is between tables, which are syntactically evident in queries: the table a query creates is dependent on the tables it references.
In contrast, current imperative practice distributes dependencies across different functions, each with custom logic and bookkeeping formats that require additional effort to navigate and make sense of.

%%%%%%%%%%%%%%%%%%%%%%%%%%%%%
\noindent\emph{\textbf{Hard Mental Operations} (high demand on cognitive resources)}.  There are two potentially challenging programming tasks in \sys.
% one
One challenge is debugging in SQL~\cite{gathani2020debugging}.
% the error could manifest downstream from the source
% not as composable compared to dataflows---not as incremental.
Consider the case where the developer is debugging a view \emph{O} which involves both \emph{V1} and \emph{V2} views; they need to inspect \emph{both} of the views to locate the error.
% instead of pin-pointing the line of code directly.
% erroneous view definition
To address this challenge, we built \emph{view level constraints}, similar to SQL table constraints~\cite{ramakrishnan2000database}, so that developers could make assertions on intermediate queries. For instance, if \emph{O} is unexpectedly empty, the developer could assert \smcode{V1 NOT EMPTY} and \smcode{V2 NOT EMPTY} respectively to pin-point the error as they arise.
% two (i.e., no side-effects)
Another challenge is not being able to mutate state.  It could be challenging to define the state of the visualization with only raw events, especially when the logic is more complex (e.g., \emph{undo-redo}).
To help, we took a page from the construct of \emph{state programs} in \emph{relational transducers}~\cite{abiteboul2000relational}, which allow developers to maintain derived state by inserting values into tables after events.
% iteratively work with history
% Lastly, we can also implement methods like static analysis to help developers code and debug, as is commonly seen in SQL IDEs.
% Even with the added language features, we would still expect some learning curve.  However, we believe that there is value in creating new abstractions if it eventually helps the developer.

%%%%%%%%%%%%%%%%%%%%%%%%%%%%%
\noindent\emph{\textbf{Diffuseness} (verbosity of language)}. 
% The somewhat cumbersome nature of relational algebra can also make \sys specifications more verbose than related declarative approaches (e.g., Vega or Vega-Lite).
The current \sys syntax hews close to SQL, and as such 
does not have syntactic conveniences one might like for
visualization (e.g., \emph{binning}~\cite{kraska2018northstar}).
% Kraska et al. have argued that SQL is cumbersome for some visualization operations (.
% We also agree that relational algebra could be cumbersome given its very small set of operators. It may make some \sys specifications more verbose. 
Some aspects of \sys's current verbosity can be alleviated by introducing syntactic sugar for common operations.  Through our own experience working with \sys and analyzing code snippets, we identified the most common programming patterns and implemented a handful of syntactic sugars. For instance, \dcode{LATEST} selects the most recent event (i.e. row(s) with the highest timestep). Similarly, the \emph{default} asynchrony policy for output views over distributed data creates an event table for the developer and selects the response for the most recent interaction. 
We provide additional details in the supplement.

\del{In sum, \sys introduces some premature commitment and hard mental operations. However, we believe these are outweighed by the \replace{decrease in viscosity}{fluidity}, and more explicit dependencies, role-expressiveness, and consistency.}

\section{Conclusion and Future Work}

% summarize contribution
By adapting two key ideas from distributed systems programming\,---\,immutable events and logical constraints\,---\,\sys contributes a substantive step towards declarative programming over distributed data and asynchronous events for interactive visualization.
Through examples, we demonstrate that developers can use \sys to declaratively specify a variety of emerging interactive visualization use cases\del{, ranging from working with remote data to visualizing interaction history}.
And, to assess the challenge that \sys's use of a relational language poses to developers, we conducted a heuristic evaluating using the Cognitive Dimensions of Notation~\cite{blackwell2001cognitive}.
We find that although \sys introduces premature commitment and possible hard mental operations, these disadvantages are outweighed by a decrease in viscosity \replace{for working with data of various sizes and changing the designs appropriately, and the increase in consistency between specifying operations over distributed data}{when working with distributed data and asynchronous events}.
Moreover, as our performance benchmarks suggest, this declarative model allows \sys to reason about the specification and optimize the execution plan.

As data grows in size and computation grows in complexity, optimizing the performance of interactive visualization application is a hot topic.
\sys's unique middleware architecture that spans the local and remotes allows for a number of research opportunities.
% one
To start, operator-level materialized-view maintenance techniques~\cite{chirkova2012materialized} can make the frontend database even faster.
\emph{Federated} databases that optimize globally across multiple databases~\cite{deshpande2002decoupled} can help us optimize data exchange between the local database and remote database.
% two
Another possibility is to automatically parallelize query evaluation~\cite{ramakrishnan2000database} across multiple threads of computation, e.g. multiple WebWorkers in a browser. 
% three
Finally, we can enhance the performance of each timestep with ``garbage collection'' by removing rows that are no longer in use from logs.  This pattern is common in many areas, such as in  replicated database systems~\cite{sarin1987discarding}, multi-version concurrency control~\cite{bernstein1981concurrency} and distributed systems programming~\cite{conway2014edelweiss}.

\add{In terms of usability, having SQL as the host language has both advantages and disadvantages (discussed in Section~\ref{sec:cog}).
To reduce the disadvantages of expressibility, future iterations of \diel may benefit from using a syntax closer to dataframe libraries like \smcode{pandas} that are better integrated with JavaScript~\cite{wu2020dataframe}.}
%   We could also explore integration with Vega and Vega-Lite to directly support scalability.

In terms of functionality, \sys does not yet support an important distributed use case, \emph{collaborative interactive visualizations} (Fig~\ref{fig:space_time_examples}\circled{3}).  Coordinating communication between multiple users is a classic challenge in distributed systems and CSCW~\cite{sun1998operational}.  A global order of events across multiple editors cannot be guaranteed without explicit coordination that decreases the interface's responsiveness.  
Instead, various coordination-free proposals have emerged that use more involved metadata than simple local timesteps to provide distributed consistency guarantees\del{, e.g., }~\cite{weiss2009logoot,conway2014language}.
% say how this could also help with the streaming use case
\add{Supporting change from multiple locations would also unlock the support for streaming, since a change from the remote database could then drive change on the client.}
It would be interesting to extend \sys with ideas from this work.

\sys is an open source system available at \url{https://github.com/yifanwu/diel}.
% By designing a unified abstraction over distributed data and asynchronous events, we hope to help developers prototype and explore alternative designs for emerging interactive visualizations faster.

\acknowledgments{ 
We thank Ryan Purpura and Lucie Choi for their contributions as undergraduate researchers at UC Berkeley; Michael Whittaker for his feedback on methods to perform the distributed execution.
The work is supported by NSF 1564049, 1564351, 1942659, 1845638, 2106197, 1452977, 1939945, 1940175, and CCF-1730628.
}

\bibliographystyle{abbrv-doi}

\bibliography{main}
\end{document}